\documentclass[reprint,
 superscriptaddress,
 amsmath,amssymb,
 aps,
]{revtex4-2}

\usepackage{colortbl} 

\usepackage[T1]{fontenc}
\usepackage{graphicx}
\usepackage{todonotes}
\usepackage{graphicx}\usepackage{dcolumn}\usepackage{bm}\usepackage{subcaption}
\usepackage{mwe}
\usepackage[normalem]{ulem}
\usepackage{tabularx}
\usepackage{booktabs}
\usepackage{tikz}
\usetikzlibrary{arrows,shapes,snakes,automata,backgrounds}
\usetikzlibrary {arrows.meta,graphs,shapes.misc,calc}

\usepackage{mathtools}
	\mathtoolsset{showonlyrefs=true}				\definecolor{sbblue}{rgb}{0.2823529411764706, 0.47058823529411764, 0.8156862745098039}
\definecolor{sborange}{rgb}{0.9333333333333333, 0.5215686274509804, 0.2901960784313726}
\definecolor{sbgreen}{rgb}{0.41568627450980394, 0.8, 0.39215686274509803}
\definecolor{sbred}{rgb}{0.8392156862745098, 0.37254901960784315, 0.37254901960784315}
\definecolor{sbpurple}{rgb}{0.5843137254901961, 0.4235294117647059, 0.7058823529411765}
\definecolor{sbbrown}{rgb}{0.5490196078431373, 0.3803921568627451, 0.23529411764705882}
\definecolor{sbmagenta}{rgb}{0.8627450980392157, 0.49411764705882355, 0.7529411764705882}
\definecolor{sbgray}{rgb}{0.4745098039215686, 0.4745098039215686, 0.4745098039215686}
\definecolor{sbocca}{rgb}{0.8352941176470589, 0.7333333333333333, 0.403921568627451}
\definecolor{sblightblue}{rgb}{0.5098039215686274, 0.7764705882352941, 0.8862745098039215}

\definecolor{lightgrey}{rgb}{0.9,0.9,0.9}
\definecolor{lightgreen}{RGB}{146,208,80}
\definecolor{lightblue}{RGB}{0,176,240}
\definecolor{darkgreen}{rgb}{0,0.39,0}

\definecolor{lightgreenblue}{RGB}{73,192,160}
\definecolor{VeryLightGray}{rgb}{0.9,0.9,0.9}
\definecolor{LightGray}{rgb}{0.8,0.8,0.8}
\definecolor{dlrgray}{RGB}{70,70,70}

\definecolor{dlightblue}{rgb}{0.65,0.65,0.98}
\definecolor{lightblue}{rgb}{0.85,0.85,1.0}
\definecolor{llightblue}{rgb}{0.95,0.95,0.98}
\definecolor{darkblue}{rgb}{0,0,.6}
\definecolor{darkred}{rgb}{.6,0,0}
\definecolor{lightred}{rgb}{1.0,0.75,0.75}
\definecolor{darkgreen}{rgb}{0,.6,0}
\definecolor{darkbrown}{rgb}{0.5,0.30,0.10}
\definecolor{grey}{rgb}{0.3,0.3,0.3}
\definecolor{red}{rgb}{.98,0,0}
\definecolor{hellgruen}{RGB}{146,208,80}
\definecolor{hellblau}{RGB}{0,176,240}
\definecolor{hellgruenblau}{RGB}{73,192,160}
\definecolor{VeryLightGray}{rgb}{0.9,0.9,0.9}
\definecolor{DLRgray}{RGB}{70,70,70}
\definecolor{dunkelblau}{RGB}{0,0,128}
\definecolor{dunkelgruen}{RGB}{0,128,0}
\definecolor{OvGUredMath}{RGB}{209,63,88}
\definecolor{AliceBlue}{rgb}{0.94,0.97,1}
\definecolor{BlueViolet}{rgb}{0.54,0.17,0.88}
\definecolor{CadetBlue}{rgb}{0.37,0.62,0.63}
\definecolor{CadetBlue1}{rgb}{0.59,0.96,1}
\definecolor{CadetBlue2}{rgb}{0.55,0.89,0.93}
\definecolor{CadetBlue3}{rgb}{0.48,0.77,0.8}
\definecolor{CadetBlue4}{rgb}{0.32,0.52,0.54}
\definecolor{CornflowerBlue}{rgb}{0.39,0.58,0.93}
\definecolor{DarkSlateBlue}{rgb}{0.28,0.24,0.54}
\definecolor{DarkTurquoise}{rgb}{0,0.8,0.82}
\definecolor{DeepSkyBlue}{rgb}{0,0.75,1}
\definecolor{DeepSkyBlue1}{rgb}{0,0.75,1}
\definecolor{DeepSkyBlue2}{rgb}{0,0.7,0.93}
\definecolor{DeepSkyBlue3}{rgb}{0,0.6,0.8}
\definecolor{DeepSkyBlue4}{rgb}{0,0.41,0.54}
\definecolor{DodgerBlue}{rgb}{0.12,0.56,1}
\definecolor{DodgerBlue1}{rgb}{0.12,0.56,1}
\definecolor{DodgerBlue2}{rgb}{0.11,0.52,0.93}
\definecolor{DodgerBlue3}{rgb}{0.09,0.45,0.8}
\definecolor{DodgerBlue4}{rgb}{0.06,0.3,0.54}
\definecolor{LightBlue}{rgb}{0.68,0.84,0.9}
\definecolor{LightBlue1}{rgb}{0.75,0.93,1}
\definecolor{LightBlue2}{rgb}{0.7,0.87,0.93}
\definecolor{LightBlue3}{rgb}{0.6,0.75,0.8}
\definecolor{LightBlue4}{rgb}{0.41,0.51,0.54}
\definecolor{LightCyan}{rgb}{0.88,1,1}
\definecolor{LightCyan1}{rgb}{0.88,1,1}
\definecolor{LightCyan2}{rgb}{0.82,0.93,0.93}
\definecolor{LightCyan3}{rgb}{0.7,0.8,0.8}
\definecolor{LightCyan4}{rgb}{0.48,0.54,0.54}
\definecolor{LightSkyBlue}{rgb}{0.53,0.8,0.98}
\definecolor{LightSkyBlue1}{rgb}{0.69,0.88,1}
\definecolor{LightSkyBlue2}{rgb}{0.64,0.82,0.93}
\definecolor{LightSkyBlue3}{rgb}{0.55,0.71,0.8}
\definecolor{LightSkyBlue4}{rgb}{0.38,0.48,0.54}
\definecolor{LightSlateBlue}{rgb}{0.52,0.44,1}
\definecolor{LightSteelBlue}{rgb}{0.69,0.77,0.87}
\definecolor{LightSteelBlue1}{rgb}{0.79,0.88,1}
\definecolor{LightSteelBlue2}{rgb}{0.73,0.82,0.93}
\definecolor{LightSteelBlue3}{rgb}{0.63,0.71,0.8}
\definecolor{LightSteelBlue4}{rgb}{0.43,0.48,0.54}
\definecolor{MediumAquamarine}{rgb}{0.4,0.8,0.66}
\definecolor{MediumBlue}{rgb}{0,0,0.8}
\definecolor{MediumSlateBlue}{rgb}{0.48,0.41,0.93}
\definecolor{MediumTurquoise}{rgb}{0.28,0.82,0.8}
\definecolor{MidnightBlue}{rgb}{0.1,0.1,0.44}
\definecolor{NavyBlue}{rgb}{0,0,0.5}
\definecolor{PaleTurquoise}{rgb}{0.68,0.93,0.93}
\definecolor{PaleTurquoise1}{rgb}{0.73,1,1}
\definecolor{PaleTurquoise2}{rgb}{0.68,0.93,0.93}
\definecolor{PaleTurquoise3}{rgb}{0.59,0.8,0.8}
\definecolor{PaleTurquoise4}{rgb}{0.4,0.54,0.54}
\definecolor{PowderBlue}{rgb}{0.69,0.88,0.9}
\definecolor{RoyalBlue}{rgb}{0.25,0.41,0.88}
\definecolor{RoyalBlue1}{rgb}{0.28,0.46,1}
\definecolor{RoyalBlue2}{rgb}{0.26,0.43,0.93}
\definecolor{RoyalBlue3}{rgb}{0.23,0.37,0.8}
\definecolor{RoyalBlue4}{rgb}{0.15,0.25,0.54}
\definecolor{SkyBlue}{rgb}{0.53,0.8,0.92}
\definecolor{SkyBlue1}{rgb}{0.53,0.8,1}
\definecolor{SkyBlue2}{rgb}{0.49,0.75,0.93}
\definecolor{SkyBlue3}{rgb}{0.42,0.65,0.8}
\definecolor{SkyBlue4}{rgb}{0.29,0.44,0.54}
\definecolor{SlateBlue}{rgb}{0.41,0.35,0.8}
\definecolor{SlateBlue1}{rgb}{0.51,0.43,1}
\definecolor{SlateBlue2}{rgb}{0.48,0.4,0.93}
\definecolor{SlateBlue3}{rgb}{0.41,0.35,0.8}
\definecolor{SlateBlue4}{rgb}{0.28,0.23,0.54}
\definecolor{SteelBlue}{rgb}{0.27,0.51,0.7}
\definecolor{SteelBlue1}{rgb}{0.39,0.72,1}
\definecolor{SteelBlue2}{rgb}{0.36,0.67,0.93}
\definecolor{SteelBlue3}{rgb}{0.31,0.58,0.8}
\definecolor{SteelBlue4}{rgb}{0.21,0.39,0.54}
\definecolor{aquamarine}{rgb}{0.5,1,0.83}
\definecolor{aquamarine1}{rgb}{0.5,1,0.83}
\definecolor{aquamarine2}{rgb}{0.46,0.93,0.77}
\definecolor{aquamarine3}{rgb}{0.4,0.8,0.66}
\definecolor{aquamarine4}{rgb}{0.27,0.54,0.45}
\definecolor{azure}{rgb}{0.94,1,1}
\definecolor{azure1}{rgb}{0.94,1,1}
\definecolor{azure2}{rgb}{0.88,0.93,0.93}
\definecolor{azure3}{rgb}{0.75,0.8,0.8}
\definecolor{azure4}{rgb}{0.51,0.54,0.54}
\definecolor{blue}{rgb}{0,0,1}
\definecolor{blue1}{rgb}{0,0,1}
\definecolor{blue2}{rgb}{0,0,0.93}
\definecolor{blue3}{rgb}{0,0,0.8}
\definecolor{blue4}{rgb}{0,0,0.54}
\definecolor{cyan}{rgb}{0,1,1}
\definecolor{cyan1}{rgb}{0,1,1}
\definecolor{cyan2}{rgb}{0,0.93,0.93}
\definecolor{cyan3}{rgb}{0,0.8,0.8}
\definecolor{cyan4}{rgb}{0,0.54,0.54}
\definecolor{navy}{rgb}{0,0,0.5}
\definecolor{turquoise}{rgb}{0.25,0.88,0.81}
\definecolor{turquoise1}{rgb}{0,0.96,1}
\definecolor{turquoise2}{rgb}{0,0.89,0.93}
\definecolor{turquoise3}{rgb}{0,0.77,0.8}
\definecolor{turquoise4}{rgb}{0,0.52,0.54}
\definecolor{RosyBrown}{rgb}{0.73,0.56,0.56}
\definecolor{RosyBrown1}{rgb}{1,0.75,0.75}
\definecolor{RosyBrown2}{rgb}{0.93,0.7,0.7}
\definecolor{RosyBrown3}{rgb}{0.8,0.61,0.61}
\definecolor{RosyBrown4}{rgb}{0.54,0.41,0.41}
\definecolor{SaddleBrown}{rgb}{0.54,0.27,0.07}
\definecolor{SandyBrown}{rgb}{0.95,0.64,0.38}
\definecolor{beige}{rgb}{0.96,0.96,0.86}
\definecolor{brown}{rgb}{0.64,0.16,0.16}
\definecolor{brown1}{rgb}{1,0.25,0.25}
\definecolor{brown2}{rgb}{0.93,0.23,0.23}
\definecolor{brown3}{rgb}{0.8,0.2,0.2}
\definecolor{brown4}{rgb}{0.54,0.14,0.14}
\definecolor{burlywood}{rgb}{0.87,0.72,0.53}
\definecolor{burlywood1}{rgb}{1,0.82,0.61}
\definecolor{burlywood2}{rgb}{0.93,0.77,0.57}
\definecolor{burlywood3}{rgb}{0.8,0.66,0.49}
\definecolor{burlywood4}{rgb}{0.54,0.45,0.33}
\definecolor{chocolate}{rgb}{0.82,0.41,0.12}
\definecolor{chocolate1}{rgb}{1,0.5,0.14}
\definecolor{chocolate2}{rgb}{0.93,0.46,0.13}
\definecolor{chocolate3}{rgb}{0.8,0.4,0.11}
\definecolor{chocolate4}{rgb}{0.54,0.27,0.07}
\definecolor{peru}{rgb}{0.8,0.52,0.25}
\definecolor{tan}{rgb}{0.82,0.7,0.55}
\definecolor{tan1}{rgb}{1,0.64,0.31}
\definecolor{tan2}{rgb}{0.93,0.6,0.29}
\definecolor{tan3}{rgb}{0.8,0.52,0.25}
\definecolor{tan4}{rgb}{0.54,0.35,0.17}
\definecolor{DarkSlateGray}{rgb}{0.18,0.31,0.31}
\definecolor{DarkSlateGray1}{rgb}{0.59,1,1}
\definecolor{DarkSlateGray2}{rgb}{0.55,0.93,0.93}
\definecolor{DarkSlateGray3}{rgb}{0.47,0.8,0.8}
\definecolor{DarkSlateGray4}{rgb}{0.32,0.54,0.54}
\definecolor{DarkSlateGrey}{rgb}{0.18,0.31,0.31}
\definecolor{DimGray}{rgb}{0.41,0.41,0.41}
\definecolor{DimGrey}{rgb}{0.41,0.41,0.41}
\definecolor{LightGray}{rgb}{0.82,0.82,0.82}
\definecolor{LightGrey}{rgb}{0.82,0.82,0.82}
\definecolor{LightSlateGray}{rgb}{0.46,0.53,0.6}
\definecolor{LightSlateGrey}{rgb}{0.46,0.53,0.6}
\definecolor{SlateGray}{rgb}{0.44,0.5,0.56}
\definecolor{SlateGray1}{rgb}{0.77,0.88,1}
\definecolor{SlateGray2}{rgb}{0.72,0.82,0.93}
\definecolor{SlateGray3}{rgb}{0.62,0.71,0.8}
\definecolor{SlateGray4}{rgb}{0.42,0.48,0.54}
\definecolor{SlateGrey}{rgb}{0.44,0.5,0.56}
\definecolor{gray}{rgb}{0.74,0.74,0.74}
\definecolor{gray0}{rgb}{0,0,0}
\definecolor{gray1}{rgb}{0.01,0.01,0.01}
\definecolor{gray10}{rgb}{0.1,0.1,0.1}
\definecolor{DarkGreen}{rgb}{0,0.39,0}
\definecolor{DarkKhaki}{rgb}{0.74,0.71,0.42}
\definecolor{DarkOliveGreen}{rgb}{0.33,0.42,0.18}
\definecolor{DarkOliveGreen1}{rgb}{0.79,1,0.44}
\definecolor{DarkOliveGreen2}{rgb}{0.73,0.93,0.41}
\definecolor{DarkOliveGreen3}{rgb}{0.63,0.8,0.35}
\definecolor{DarkOliveGreen4}{rgb}{0.43,0.54,0.24}
\definecolor{DarkSeaGreen}{rgb}{0.56,0.73,0.56}
\definecolor{DarkSeaGreen1}{rgb}{0.75,1,0.75}
\definecolor{DarkSeaGreen2}{rgb}{0.7,0.93,0.7}
\definecolor{DarkSeaGreen3}{rgb}{0.61,0.8,0.61}
\definecolor{DarkSeaGreen4}{rgb}{0.41,0.54,0.41}
\definecolor{ForestGreen}{rgb}{0.13,0.54,0.13}
\definecolor{GreenYellow}{rgb}{0.68,1,0.18}
\definecolor{LawnGreen}{rgb}{0.48,0.98,0}
\definecolor{LightSeaGreen}{rgb}{0.13,0.7,0.66}
\definecolor{LimeGreen}{rgb}{0.2,0.8,0.2}
\definecolor{MediumSeaGreen}{rgb}{0.23,0.7,0.44}
\definecolor{MediumSpringGreen}{rgb}{0,0.98,0.6}
\definecolor{MintCream}{rgb}{0.96,1,0.98}
\definecolor{OliveDrab}{rgb}{0.42,0.55,0.14}
\definecolor{OliveDrab1}{rgb}{0.75,1,0.24}
\definecolor{OliveDrab2}{rgb}{0.7,0.93,0.23}
\definecolor{OliveDrab3}{rgb}{0.6,0.8,0.2}
\definecolor{OliveDrab4}{rgb}{0.41,0.54,0.13}
\definecolor{PaleGreen}{rgb}{0.59,0.98,0.59}
\definecolor{PaleGreen1}{rgb}{0.6,1,0.6}
\definecolor{PaleGreen2}{rgb}{0.56,0.93,0.56}
\definecolor{PaleGreen3}{rgb}{0.48,0.8,0.48}
\definecolor{PaleGreen4}{rgb}{0.33,0.54,0.33}
\definecolor{SeaGreen}{rgb}{0.18,0.54,0.34}
\definecolor{SeaGreen1}{rgb}{0.33,1,0.62}
\definecolor{SeaGreen2}{rgb}{0.3,0.93,0.58}
\definecolor{SeaGreen3}{rgb}{0.26,0.8,0.5}
\definecolor{SeaGreen4}{rgb}{0.18,0.54,0.34}
\definecolor{SpringGreen}{rgb}{0,1,0.5}
\definecolor{SpringGreen1}{rgb}{0,1,0.5}
\definecolor{SpringGreen2}{rgb}{0,0.93,0.46}
\definecolor{SpringGreen3}{rgb}{0,0.8,0.4}
\definecolor{SpringGreen4}{rgb}{0,0.54,0.27}
\definecolor{YellowGreen}{rgb}{0.6,0.8,0.2}
\definecolor{chartreuse}{rgb}{0.5,1,0}
\definecolor{chartreuse1}{rgb}{0.5,1,0}
\definecolor{chartreuse2}{rgb}{0.46,0.93,0}
\definecolor{chartreuse3}{rgb}{0.4,0.8,0}
\definecolor{chartreuse4}{rgb}{0.27,0.54,0}
\definecolor{green}{rgb}{0,1,0}
\definecolor{green1}{rgb}{0,1,0}
\definecolor{green2}{rgb}{0,0.93,0}
\definecolor{green3}{rgb}{0,0.8,0}
\definecolor{green4}{rgb}{0,0.54,0}
\definecolor{khaki}{rgb}{0.94,0.9,0.55}
\definecolor{khaki1}{rgb}{1,0.96,0.56}
\definecolor{khaki2}{rgb}{0.93,0.9,0.52}
\definecolor{khaki3}{rgb}{0.8,0.77,0.45}
\definecolor{khaki4}{rgb}{0.54,0.52,0.3}
\definecolor{DarkOrange}{rgb}{1,0.55,0}
\definecolor{DarkOrange1}{rgb}{1,0.5,0}
\definecolor{DarkOrange2}{rgb}{0.93,0.46,0}
\definecolor{DarkOrange3}{rgb}{0.8,0.4,0}
\definecolor{DarkOrange4}{rgb}{0.54,0.27,0}
\definecolor{DarkSalmon}{rgb}{0.91,0.59,0.48}
\definecolor{LightCoral}{rgb}{0.94,0.5,0.5}
\definecolor{LightSalmon}{rgb}{1,0.63,0.48}
\definecolor{LightSalmon1}{rgb}{1,0.63,0.48}
\definecolor{LightSalmon2}{rgb}{0.93,0.58,0.45}
\definecolor{LightSalmon3}{rgb}{0.8,0.5,0.38}
\definecolor{LightSalmon4}{rgb}{0.54,0.34,0.26}
\definecolor{PeachPuff}{rgb}{1,0.85,0.72}
\definecolor{PeachPuff1}{rgb}{1,0.85,0.72}
\definecolor{PeachPuff2}{rgb}{0.93,0.79,0.68}
\definecolor{PeachPuff3}{rgb}{0.8,0.68,0.58}
\definecolor{PeachPuff4}{rgb}{0.54,0.46,0.39}
\definecolor{bisque}{rgb}{1,0.89,0.77}
\definecolor{bisque1}{rgb}{1,0.89,0.77}
\definecolor{bisque2}{rgb}{0.93,0.83,0.71}
\definecolor{bisque3}{rgb}{0.8,0.71,0.62}
\definecolor{bisque4}{rgb}{0.54,0.49,0.42}
\definecolor{coral}{rgb}{1,0.5,0.31}
\definecolor{coral1}{rgb}{1,0.45,0.34}
\definecolor{coral2}{rgb}{0.93,0.41,0.31}
\definecolor{coral3}{rgb}{0.8,0.36,0.27}
\definecolor{coral4}{rgb}{0.54,0.24,0.18}
\definecolor{honeydew}{rgb}{0.94,1,0.94}
\definecolor{honeydew1}{rgb}{0.94,1,0.94}
\definecolor{honeydew2}{rgb}{0.88,0.93,0.88}
\definecolor{honeydew3}{rgb}{0.75,0.8,0.75}
\definecolor{honeydew4}{rgb}{0.51,0.54,0.51}
\definecolor{orange}{rgb}{1,0.64,0}
\definecolor{orange1}{rgb}{1,0.64,0}
\definecolor{orange2}{rgb}{0.93,0.6,0}
\definecolor{orange3}{rgb}{0.8,0.52,0}
\definecolor{orange4}{rgb}{0.54,0.35,0}
\definecolor{salmon}{rgb}{0.98,0.5,0.45}
\definecolor{salmon1}{rgb}{1,0.55,0.41}
\definecolor{salmon2}{rgb}{0.93,0.51,0.38}
\definecolor{salmon3}{rgb}{0.8,0.44,0.33}
\definecolor{salmon4}{rgb}{0.54,0.3,0.22}
\definecolor{sienna}{rgb}{0.63,0.32,0.18}
\definecolor{sienna1}{rgb}{1,0.51,0.28}
\definecolor{sienna2}{rgb}{0.93,0.47,0.26}
\definecolor{sienna3}{rgb}{0.8,0.41,0.22}
\definecolor{sienna4}{rgb}{0.54,0.28,0.15}
\definecolor{DeepPink}{rgb}{1,0.08,0.57}
\definecolor{DeepPink1}{rgb}{1,0.08,0.57}
\definecolor{DeepPink2}{rgb}{0.93,0.07,0.54}
\definecolor{DeepPink3}{rgb}{0.8,0.06,0.46}
\definecolor{DeepPink4}{rgb}{0.54,0.04,0.31}
\definecolor{HotPink}{rgb}{1,0.41,0.7}
\definecolor{HotPink1}{rgb}{1,0.43,0.7}
\definecolor{HotPink2}{rgb}{0.93,0.41,0.65}
\definecolor{HotPink3}{rgb}{0.8,0.38,0.56}
\definecolor{HotPink4}{rgb}{0.54,0.23,0.38}
\definecolor{IndianRed}{rgb}{0.8,0.36,0.36}
\definecolor{IndianRed1}{rgb}{1,0.41,0.41}
\definecolor{IndianRed2}{rgb}{0.93,0.39,0.39}
\definecolor{IndianRed3}{rgb}{0.8,0.33,0.33}
\definecolor{IndianRed4}{rgb}{0.54,0.23,0.23}
\definecolor{LightPink}{rgb}{1,0.71,0.75}
\definecolor{LightPink1}{rgb}{1,0.68,0.72}
\definecolor{LightPink2}{rgb}{0.93,0.63,0.68}
\definecolor{LightPink3}{rgb}{0.8,0.55,0.58}
\definecolor{LightPink4}{rgb}{0.54,0.37,0.39}
\definecolor{MediumVioletRed}{rgb}{0.78,0.08,0.52}
\definecolor{MistyRose}{rgb}{1,0.89,0.88}
\definecolor{MistyRose1}{rgb}{1,0.89,0.88}
\definecolor{MistyRose2}{rgb}{0.93,0.83,0.82}
\definecolor{MistyRose3}{rgb}{0.8,0.71,0.71}
\definecolor{MistyRose4}{rgb}{0.54,0.49,0.48}
\definecolor{OrangeRed}{rgb}{1,0.27,0}
\definecolor{OrangeRed1}{rgb}{1,0.27,0}
\definecolor{OrangeRed2}{rgb}{0.93,0.25,0}
\definecolor{OrangeRed3}{rgb}{0.8,0.21,0}
\definecolor{OrangeRed4}{rgb}{0.54,0.14,0}
\definecolor{PaleVioletRed}{rgb}{0.86,0.44,0.57}
\definecolor{PaleVioletRed1}{rgb}{1,0.51,0.67}
\definecolor{PaleVioletRed2}{rgb}{0.93,0.47,0.62}
\definecolor{PaleVioletRed3}{rgb}{0.8,0.41,0.54}
\definecolor{PaleVioletRed4}{rgb}{0.54,0.28,0.36}
\definecolor{VioletRed}{rgb}{0.81,0.13,0.56}
\definecolor{VioletRed1}{rgb}{1,0.24,0.59}
\definecolor{VioletRed2}{rgb}{0.93,0.23,0.55}
\definecolor{VioletRed3}{rgb}{0.8,0.2,0.47}
\definecolor{VioletRed4}{rgb}{0.54,0.13,0.32}
\definecolor{firebrick}{rgb}{0.7,0.13,0.13}
\definecolor{firebrick1}{rgb}{1,0.19,0.19}
\definecolor{firebrick2}{rgb}{0.93,0.17,0.17}
\definecolor{firebrick3}{rgb}{0.8,0.15,0.15}
\definecolor{firebrick4}{rgb}{0.54,0.1,0.1}
\definecolor{pink}{rgb}{1,0.75,0.79}
\definecolor{pink1}{rgb}{1,0.71,0.77}
\definecolor{pink2}{rgb}{0.93,0.66,0.72}
\definecolor{pink3}{rgb}{0.8,0.57,0.62}
\definecolor{pink4}{rgb}{0.54,0.39,0.42}
\definecolor{red}{rgb}{1,0,0}
\definecolor{red1}{rgb}{1,0,0}
\definecolor{red2}{rgb}{0.93,0,0}
\definecolor{red3}{rgb}{0.8,0,0}
\definecolor{red4}{rgb}{0.54,0,0}
\definecolor{tomato}{rgb}{1,0.39,0.28}
\definecolor{tomato1}{rgb}{1,0.39,0.28}
\definecolor{tomato2}{rgb}{0.93,0.36,0.26}
\definecolor{tomato3}{rgb}{0.8,0.31,0.22}
\definecolor{tomato4}{rgb}{0.54,0.21,0.15}
\definecolor{DarkOrchid}{rgb}{0.6,0.2,0.8}
\definecolor{DarkOrchid1}{rgb}{0.75,0.24,1}
\definecolor{DarkOrchid2}{rgb}{0.7,0.23,0.93}
\definecolor{DarkOrchid3}{rgb}{0.6,0.2,0.8}
\definecolor{DarkOrchid4}{rgb}{0.41,0.13,0.54}
\definecolor{DarkViolet}{rgb}{0.58,0,0.82}
\definecolor{LavenderBlush}{rgb}{1,0.94,0.96}
\definecolor{LavenderBlush1}{rgb}{1,0.94,0.96}
\definecolor{LavenderBlush2}{rgb}{0.93,0.88,0.89}
\definecolor{LavenderBlush3}{rgb}{0.8,0.75,0.77}
\definecolor{LavenderBlush4}{rgb}{0.54,0.51,0.52}
\definecolor{MediumOrchid}{rgb}{0.73,0.33,0.82}
\definecolor{MediumOrchid1}{rgb}{0.88,0.4,1}
\definecolor{MediumOrchid2}{rgb}{0.82,0.37,0.93}
\definecolor{MediumOrchid3}{rgb}{0.7,0.32,0.8}
\definecolor{MediumOrchid4}{rgb}{0.48,0.21,0.54}
\definecolor{MediumPurple}{rgb}{0.57,0.44,0.86}
\definecolor{MediumPurple1}{rgb}{0.67,0.51,1}
\definecolor{MediumPurple2}{rgb}{0.62,0.47,0.93}
\definecolor{MediumPurple3}{rgb}{0.54,0.41,0.8}
\definecolor{MediumPurple4}{rgb}{0.36,0.28,0.54}
\definecolor{lavender}{rgb}{0.9,0.9,0.98}
\definecolor{magenta}{rgb}{1,0,1}
\definecolor{magenta1}{rgb}{1,0,1}
\definecolor{magenta2}{rgb}{0.93,0,0.93}
\definecolor{magenta3}{rgb}{0.8,0,0.8}
\definecolor{magenta4}{rgb}{0.54,0,0.54}
\definecolor{maroon}{rgb}{0.69,0.19,0.38}
\definecolor{maroon1}{rgb}{1,0.2,0.7}
\definecolor{maroon2}{rgb}{0.93,0.19,0.65}
\definecolor{maroon3}{rgb}{0.8,0.16,0.56}
\definecolor{maroon4}{rgb}{0.54,0.11,0.38}
\definecolor{orchid}{rgb}{0.85,0.44,0.84}
\definecolor{orchid1}{rgb}{1,0.51,0.98}
\definecolor{orchid2}{rgb}{0.93,0.48,0.91}
\definecolor{orchid3}{rgb}{0.8,0.41,0.79}
\definecolor{orchid4}{rgb}{0.54,0.28,0.54}
\definecolor{plum}{rgb}{0.86,0.63,0.86}
\definecolor{plum1}{rgb}{1,0.73,1}
\definecolor{plum2}{rgb}{0.93,0.68,0.93}
\definecolor{plum3}{rgb}{0.8,0.59,0.8}
\definecolor{plum4}{rgb}{0.54,0.4,0.54}
\definecolor{purple}{rgb}{0.63,0.13,0.94}
\definecolor{purple1}{rgb}{0.61,0.19,1}
\definecolor{purple2}{rgb}{0.57,0.17,0.93}
\definecolor{purple3}{rgb}{0.49,0.15,0.8}
\definecolor{purple4}{rgb}{0.33,0.1,0.54}
\definecolor{thistle}{rgb}{0.84,0.75,0.84}
\definecolor{thistle1}{rgb}{1,0.88,1}
\definecolor{thistle2}{rgb}{0.93,0.82,0.93}
\definecolor{thistle3}{rgb}{0.8,0.71,0.8}
\definecolor{thistle4}{rgb}{0.54,0.48,0.54}
\definecolor{violet}{rgb}{0.93,0.51,0.93}
\definecolor{AntiqueWhite}{rgb}{0.98,0.92,0.84}
\definecolor{AntiqueWhite1}{rgb}{1,0.93,0.86}
\definecolor{AntiqueWhite2}{rgb}{0.93,0.87,0.8}
\definecolor{AntiqueWhite3}{rgb}{0.8,0.75,0.69}
\definecolor{AntiqueWhite4}{rgb}{0.54,0.51,0.47}
\definecolor{FloralWhite}{rgb}{1,0.98,0.94}
\definecolor{GhostWhite}{rgb}{0.97,0.97,1}
\definecolor{NavajoWhite}{rgb}{1,0.87,0.68}
\definecolor{NavajoWhite1}{rgb}{1,0.87,0.68}
\definecolor{NavajoWhite2}{rgb}{0.93,0.81,0.63}
\definecolor{NavajoWhite3}{rgb}{0.8,0.7,0.54}
\definecolor{NavajoWhite4}{rgb}{0.54,0.47,0.37}
\definecolor{OldLace}{rgb}{0.99,0.96,0.9}
\definecolor{WhiteSmoke}{rgb}{0.96,0.96,0.96}
\definecolor{gainsboro}{rgb}{0.86,0.86,0.86}
\definecolor{ivory}{rgb}{1,1,0.94}
\definecolor{ivory1}{rgb}{1,1,0.94}
\definecolor{ivory2}{rgb}{0.93,0.93,0.88}
\definecolor{ivory3}{rgb}{0.8,0.8,0.75}
\definecolor{ivory4}{rgb}{0.54,0.54,0.51}
\definecolor{linen}{rgb}{0.98,0.94,0.9}
\definecolor{seashell}{rgb}{1,0.96,0.93}
\definecolor{seashell1}{rgb}{1,0.96,0.93}
\definecolor{seashell2}{rgb}{0.93,0.89,0.87}
\definecolor{seashell3}{rgb}{0.8,0.77,0.75}
\definecolor{seashell4}{rgb}{0.54,0.52,0.51}
\definecolor{snow}{rgb}{1,0.98,0.98}
\definecolor{snow1}{rgb}{1,0.98,0.98}
\definecolor{snow2}{rgb}{0.93,0.91,0.91}
\definecolor{snow3}{rgb}{0.8,0.79,0.79}
\definecolor{snow4}{rgb}{0.54,0.54,0.54}
\definecolor{wheat}{rgb}{0.96,0.87,0.7}
\definecolor{wheat1}{rgb}{1,0.9,0.73}
\definecolor{wheat2}{rgb}{0.93,0.84,0.68}
\definecolor{wheat3}{rgb}{0.8,0.73,0.59}
\definecolor{wheat4}{rgb}{0.54,0.49,0.4}
\definecolor{white}{rgb}{1,1,1}
\definecolor{BlanchedAlmond}{rgb}{1,0.92,0.8}
\definecolor{DarkGoldenrod}{rgb}{0.72,0.52,0.04}
\definecolor{DarkGoldenrod1}{rgb}{1,0.72,0.06}
\definecolor{DarkGoldenrod2}{rgb}{0.93,0.68,0.05}
\definecolor{DarkGoldenrod3}{rgb}{0.8,0.58,0.05}
\definecolor{DarkGoldenrod4}{rgb}{0.54,0.39,0.03}
\definecolor{LemonChiffon}{rgb}{1,0.98,0.8}
\definecolor{LemonChiffon1}{rgb}{1,0.98,0.8}
\definecolor{LemonChiffon2}{rgb}{0.93,0.91,0.75}
\definecolor{LemonChiffon3}{rgb}{0.8,0.79,0.64}
\definecolor{LemonChiffon4}{rgb}{0.54,0.54,0.44}
\definecolor{LightGoldenrod}{rgb}{0.93,0.86,0.51}
\definecolor{LightGoldenrod1}{rgb}{1,0.92,0.54}
\definecolor{LightGoldenrod2}{rgb}{0.93,0.86,0.51}
\definecolor{LightGoldenrod3}{rgb}{0.8,0.74,0.44}
\definecolor{LightGoldenrod4}{rgb}{0.54,0.5,0.3}
\definecolor{LightGoldenrodYellow}{rgb}{0.98,0.98,0.82}
\definecolor{LightYellow}{rgb}{1,1,0.88}
\definecolor{LightYellow1}{rgb}{1,1,0.88}
\definecolor{LightYellow2}{rgb}{0.93,0.93,0.82}
\definecolor{LightYellow3}{rgb}{0.8,0.8,0.7}
\definecolor{LightYellow4}{rgb}{0.54,0.54,0.48}
\definecolor{PaleGoldenrod}{rgb}{0.93,0.91,0.66}
\definecolor{PapayaWhip}{rgb}{1,0.93,0.83}
\definecolor{cornsilk}{rgb}{1,0.97,0.86}
\definecolor{cornsilk1}{rgb}{1,0.97,0.86}
\definecolor{cornsilk2}{rgb}{0.93,0.91,0.8}
\definecolor{cornsilk3}{rgb}{0.8,0.78,0.69}
\definecolor{cornsilk4}{rgb}{0.54,0.53,0.47}
\definecolor{gold}{rgb}{1,0.84,0}
\definecolor{gold1}{rgb}{1,0.84,0}
\definecolor{gold2}{rgb}{0.93,0.79,0}
\definecolor{gold3}{rgb}{0.8,0.68,0}
\definecolor{gold4}{rgb}{0.54,0.46,0}
\definecolor{goldenrod}{rgb}{0.85,0.64,0.13}
\definecolor{goldenrod1}{rgb}{1,0.75,0.14}
\definecolor{goldenrod2}{rgb}{0.93,0.7,0.13}
\definecolor{goldenrod3}{rgb}{0.8,0.61,0.11}
\definecolor{goldenrod4}{rgb}{0.54,0.41,0.08}
\definecolor{moccasin}{rgb}{1,0.89,0.71}
\definecolor{yellow}{rgb}{1,1,0}
\definecolor{yellow1}{rgb}{1,1,0}
\definecolor{yellow2}{rgb}{0.93,0.93,0}
\definecolor{yellow3}{rgb}{0.8,0.8,0}
\definecolor{yellow4}{rgb}{0.54,0.54,0}

\usepackage{hyperref}
\hypersetup{colorlinks=true, linkcolor=blue, citecolor=ForestGreen, urlcolor=DarkOrchid}

\newcommand{\ket}[1]{|#1 \rangle}

\newcommand{\dep}{^\text{d}}
\newcommand{\arr}{^\text{a}}
\newcommand{\iin}{^\text{in}}
\newcommand{\out}{^\text{out}}
\newcommand{\buf}{^\text{buf}}
 
\begin{document}
\title{Performance of Domain-Wall Encoding for Quantum Annealing}

\author{Jie Chen}\affiliation{Department of Physics; Joint Quantum Centre (JQC) Durham-Newcastle \\ Durham University, South Road, Durham, DH1 3LE, UK}\author{Tobias Stollenwerk}\affiliation{German Aerospace Center (DLR), Linder H\"ohe, 51147 Cologne, Germany}
\author{Nicholas Chancellor}\affiliation{Department of Physics; Joint Quantum Centre (JQC) Durham-Newcastle \\ Durham University, South Road, Durham, DH1 3LE, UK}\date{December 2020}

\begin{abstract}
    In this paper we experimentally test the performance of the recently proposed domain-wall encoding of discrete variables from [Chancellor Quantum Sci. Technol. 4 045004] on Ising model flux qubit quantum annealers. We compare this encoding with the traditional one-hot methods and find that they outperform the one-hot encoding for three different problems at different sizes both of the problem and of the variables. From these results we conclude that the domain-wall encoding yields superior performance against a variety of metrics furthermore, we do not find a single metric by which one hot performs better. We even find that a 2000Q quantum annealer with a drastically less connected hardware graph but using the domain-wall encoding can outperform the next generation Advantage processor if that processor uses one-hot encoding.
\end{abstract}
 
\maketitle

\section{Introduction}
Quantum annealing is a subject of much recent interest, because of recent advances in both theory and experimental implementations. After the initial numerical studies which pointed to quantum annealing as a potential tool for optimization \cite{kadowaki98a}, focus was mostly on relatively simple closed systems in the adiabatic limit \cite{farhi00a,albash16a}. However, a wide variety of advances have now taken place, for example better understanding of the role noise plays \cite{Venuti16open}, more rapid quenches \cite{Morley19a,Hastings19a,Callison20a,crosson2020prospects}, and how to incorporate quantum annealing into hybrid protocols \cite{Perdomo-Ortiz11,Duan2013a,chancellor17b,Grass19a}. Experimentally this field is exciting because it allows for large scale experiments on superconducting hardware designed to solve difficult optimization problems. Proof-of-concept studies have taken place on a diverse range of topics including aerospace problems\cite{stollenwerkFGA2019,stollenwerkATM2019}, hydrology \cite{omalley18a}, radar waveform design \cite{coxson14a}, scheduling \cite{crispin13a,Venturelli15a,Tran16a} and traffic flow optimization~\cite{neukart2017,yarkoni2020}. While to our knowledge a scaling advantage for optimization has yet to be seen, signs of a potential advantage have been observed in recent quantum simulation experiments \cite{King2019b}.

The problems which these devices solve are encoded as energy minimization with respect to quadratic unconstrained binary (QUBO) Hamiltonians (aka. penalty functions) of the form
\begin{equation}
    H_{\mathrm{QUBO}}=\sum_{i,j\in\chi}Q_{ij}b_ib_j,
\end{equation}
where $b_i\in \{0,1\}$ encodes the value of qubit $i$ and $Q$ is the QUBO matrix which defines the problem\footnote{note that $b^2_i=b_i$}. 
Similarly, one could use the equivalent Ising formulation by substituting $b_i = (1 - z_i)/2$, $z_i  \in \{1,-1\}$.
The interactions are constrained to a graph $\chi$, but as long as the graph obeys certain structural constraints, arbitrary connectivity can be mapped using a technique known as minor embedding \cite{choi08a,choi10a} where variables within a graph minor are joined using strong ferromagnetic (negative) interactions. These joined variables are referred to as chains, and the strength of the interactions is referred to as chain strength. An alternative approach to mapping is to use parity constraints \cite{Lechner15a,Rocchetto16a,Leib16a,chancellor17a}, but we use minor embedding methods for this paper because they are more commonly used, and because quantum Monte-Carlo studies have suggested that this is a better method for the kind of devices we study here \cite{Albash16b}.

 Since solving a QUBO is known to be NP-hard, all other optimization problems can be mapped to them with only a polynomial overhead. One particular mapping which is common is a discrete-to-binary mapping, where discrete variables with greater than two values are mapped to binary variables. The traditional way to do this is to use a kind of constraint known as a one-hot constraint which requires that only one of a set of qubits can be in the $\ket{1}$ configuration. 

To construct a quantum algorithm to solve a QUBO problem, quantum annealing is performed, the Hamiltonian which describes this process includes Pauli $X$ terms ($X_i$) which introduce quantum mechanical (qu)bit flips
\begin{equation}
    H(A,B)=-A(t)\sum_i X_i+B(t)H_{\mathrm{QUBO}}
\end{equation}
where the protocol starts out in an equal positive superposition of all possible solutions with $\frac{A(t=0)}{B(t=0)}\gg 1$, and ends with $\frac{B(t=t_f)}{A(t=t_f)}\gg 1$. 
Note, that the first term, which is called \textit{driver}, can be replaced by other operators which do not commute with the second operator.
While we are not concerned with the detailed physics of how these devices operate in the present study, it is worth remarking that the devices we study here operate in a highly dissipative regime, where interactions with a low-temperature environment play an important role in the dynamics.

Recently it has been demonstrated in \cite{chancellor2019} that a different way of encoding discrete variables, known as domain-wall encoding, can lead to problem structures which make minor-embedding more efficiently and use fewer variables than one-hot encodings while still allow arbitrary interactions between the variables. This study was purely numerical and theoretical, and furthermore has pointed out that the qubit flips explore the solution space in fundamentally different ways for one-hot versus domain-wall encoded problems. The domain-wall encoding has found use in quantum simulations of quantum field theories \cite{Abel20a,Abel20b} and has been used in proof-of-concept experiments for using quantum fluctuations to guide searches on annealers \cite{Chancellor20a}. To our knowledge, however, there has never been a direct experimental test of the relative performance between between domain-wall and one-hot encodings. 

Since the solution space is not explored in the same way for the two encodings, it is not \emph{a priori} clear that the more efficient embedding will translate to improved problem solving abilities . The search could be less effective in a way which negates the gains from improved embedding. However when we perform the experiments on several examples we find that the domain wall encoding does indeed lead to an improvement over many different metrics. These experiments are performed on two different quantum processing units (QPUs) manufactured by D-Wave Systems Inc.~which have different allowed interaction graphs, an older, less connected generation (2000Q), and a newer more connected one (Advantage). We find that at least by some metrics, the use of the more sophisticated domain-wall encoding can make more of a difference to the ability to solve problems than the re-engineered hardware graphs. Although not the primary goal of this paper, we also compare between the two QPU architectures (and between all QPU-encoding combinations), this allows us to compare the gains from using the domain wall encoding to those attained by using a more connected architecture.

 \section{Discrete Quadratic Models}
While the native models for quantum annealers are quadratic unconstrained binary optimization problems (QUBOs), many real world problems are most naturally expressed in a way which still involves pairwise interactions between terms, these are referred to as discrete quadratic models (DQMs). 
A DQM can be described by a set of discrete variables $d_i$, $i \in [n-1]$, as well as 
arbitrary pairwise interactions between these variables. Note that the elements in $d_i$ do not need to be integers or even a number, it can be any discrete set, for example colors in a co louring problem, but are indexed in order by the index $i$.
A DQM Hamiltonian can be written as an extension of a QUBO Hamiltonian with an additional index denoting the variable value,
\begin{equation}
    H_{\mathrm{DQM}}=\sum_{i,j}\sum_{\alpha,\beta} D_{(i,j,\alpha,\beta)}{x}_{i,\alpha}{x}_{j,\beta},
\end{equation}
where 
\begin{equation}
   {x}_{i,\alpha}=\begin{cases}1 & \mathrm{variable}\,\, d_i\,\, \mathrm{takes\,\, value}\,\, \alpha \\ 0 & \mathrm{otherwise}\end{cases}
\end{equation}
and $D_{(i,j,\alpha,\beta)}$ defines the pairwise interactions between the variables. 
For the sake of simplicity, we further constrain the values of the discrete variables to consecutive integers $\alpha \in [m-1]$ in this work.  An example for such discrete variables is the colors $\alpha$ of vertex $i$ in graph coloring problems.
The extensions to arbitrary sets of discrete values is straight forward. 
See~\cite{stollenwerkATM2019} as an example.

\subsection{Domain-Wall and One-Hot Encoding}
The discrete variable $d_i$ is encoded in multiple binary variables $x_{i\alpha}$.
We now quickly review the encoding methods which are commonly used for these discrete variables. 
The traditional method is known as one-hot encoding.
Here, each qubit corresponds to one possible value of the discrete variable and a constraint which specifies that the variable only takes one value has to be imposed
\begin{equation} \label{eq:constraint}
    \forall i \, \sum_\alpha x_{i, \alpha} = 1 \, .
\end{equation}
This constraint can be enforced by adding a quadratic penalty term to the Hamiltonian
\begin{align}
    H_{\mathrm{one\,hot}}= &H_{\mathrm{DQM}} 
                           + \lambda \sum_i  \left(\sum^{m-1}_{\alpha=0} x_{i,\alpha}-1\right)^2.\label{eq:Hoh}
\end{align}
Hence, we can write the discrete variables as $d_i = \sum_{\alpha=0}^m \alpha x_{i,\alpha}$.
However, from the perspective of physically implementation on a real device, it has the undesirable property that all qubits used to encode a variable must be able to interact with all others used to encode that same variables. 

\begin{figure}[htpb]
    \centering
    \begin{tikzpicture}[qubit/.style={circle, inner sep=0pt, minimum size=5pt, fill=black},
        outer/.style={circle, draw, inner sep=0pt, minimum size=5pt, fill=gray},
        ]
        \tikzset{node distance = 26pt},
        \tikzset{label distance = 8pt},
        \node[outer, label=below:${s}_{i, -1}$] at (0, 0) (-1) {}; 
        \node[qubit, label=below:${s}_{i, 0}$,   right=of -1] (0) {};
        \node[qubit, label=below:${s}_{i, 1}$,   right=of 0] (1) {};
        \node[qubit, label=below:${s}_{i, 2}$,   right=of 1] (2) {};
        \node[right= of 2]        (dots) {$\cdots$};
        \node[qubit, label=below:${s}_{i, m-2}$,   right=of dots] (m-2) {};
        \node[outer, label=below:${s}_{i, m-1}$, right=of m-2] (m-1) {};

        \draw[]   (-1) -- node[above, midway] {$\kappa$}  (0) ;
        \draw[]    (0) -- node[above, midway] {$\kappa$}  (1) ;
        \draw[]    (1) -- node[above, midway] {$\kappa$}  (2) ;
        \draw[]    (2) -- node[above, midway] {$\kappa$}  (dots) ;
        \draw[] (dots) -- node[above, midway] {$\kappa$}  (m-2) ;
        \draw[]    (m-2) -- node[above, midway] {$\kappa$}  (m-1) ;

\begin{scope}[on background layer]
            \draw[-{Stealth}, thick] ($(-1)  + (0, 8pt)$)  --  ($(-1)  + (0, -10pt)$);
            \draw[-{Stealth}, thick] ($(0)   + (0, 8pt)$)  --  ($(0)   + (0, -10pt)$);
            \draw[-{Stealth}, thick] ($(1)   + (0, 8pt)$)  --  ($(1)   + (0, -10pt)$);
            \draw[-{Stealth}, thick] ($(2)   + (0, -8pt)$) --  ($(2)   + (0, 10pt)$);
            \draw[-{Stealth}, thick] ($(m-2) + (0, -8pt)$) --  ($(m-2) + (0, 10pt)$);
            \draw[-{Stealth}, thick] ($(m-1)   + (0, -8pt)$) --  ($(m-1)   + (0, 10pt)$);
            \draw[dashed] ($(1)   + (16pt, -25pt)$) --  ($(1)   + (16pt, 25pt)$);
        \end{scope}
    \end{tikzpicture}
    \caption{Domain wall encoding scheme. The value of the discrete variable $d_i$ is given by the position $\alpha\in [m-1]$ of the domain wall (indicated by dashed line) in this Ising chain with fixed outer spins ${s}_{i, -1} = -1$ and ${s}_{i, m-1}=1$.  }\label{fig:dwe_scheme}
\end{figure}
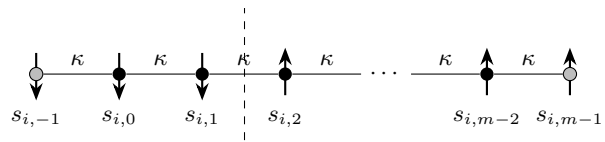

It was shown in \cite{chancellor2019} that instead using a ``domain wall'' encoding strategy can encode discrete variables with one fewer qubit per variable and does not require interactions between all qubits to implement the constraint. This work found that minor embedding was more efficient when the domain-wall encoding was used. While a full description of this encoding strategy can be be found in \cite{chancellor2019}, and Python code to implement the domain-wall encoding can be found at \cite{domain_wall_code} we review the key details here in the interest of making this manuscript self contained. 
The underlying principle of the domain wall encoding is to use the degeneracy of domain wall positions on a segment of a frustrated Ising spin chain as it is shown in Figure~\ref{fig:dwe_scheme}.
For each discrete variable $d_i$,
we have $m-1$ spin variables and two fixed spins at the edges of the chain ${s}_{i, -1} = -1$ and ${s}_{i, m-1}=1$.
The variable $d_i$ is correctly encoded if there is a single domain wall in the Ising chain.
This is enforced by the penalty Hamiltonian describing a ferromagnetic coupling between the Ising spin variables

\begin{align}
    H_{\mathrm{chain}} & = -\kappa\left(\sum^{m-2}_{\alpha=-1}{s}_{i, \alpha}{s}_{i, \alpha+1}\right) \label{eq:Hdw} 
\end{align}
where $\kappa$ is a coupling large enough to enforce a single domain wall in the ground state (cf.~\cite{chancellor2019}). A concrete example of one-hot and domain wall encodings ($m=4$ for one-hot and $m=4$ for domain-wall is provided in table \ref{tab:dw_oh}.
Then, the binary variable $x_{i, \alpha}$ depends on the values of the spin variable according to 
\begin{equation}
    x_{i, \alpha} = \frac{1}{2} \left( {s}_{i, \alpha} - {s}_{i, \alpha - 1} \right) \quad , \forall i, \alpha \in [n-1] \times [m-1] \, ,
\end{equation}
and Equation~\eqref{eq:constraint} is fulfilled.
Note that only $m-1$ variables (the inner spins) are needed to encode a variable with $m$ different values.
This is one less than in the one-hot-encoding case.
Hence, the total Hamiltonian for the DQM reads
\begin{align}
    H_{\text{domain wall}}= &H_{\mathrm{DQM}}+ H_\text{chain}
\end{align}
Note that this is a function purely of the $(m-1)n$ (\textit{inner}) spin variables $\{s_{i, \alpha} \, | \, i \in [n-1],  \alpha \in [m - 2] \}$.
The conversion back to a QUBO is straight forward.

\begin{table}[t]

\begin{tabular}{|c|c|c|}
\hline 
value & one-hot & domain-wall\tabularnewline
\hline 
\hline 
0 & N/A & 0000\tabularnewline
\hline 
1 & 1000 & 1000\tabularnewline
\hline 
2 & 0100 & 1100\tabularnewline
\hline 
3 & 0010 & 1110\tabularnewline
\hline 
4 & 0001 & 1111\tabularnewline
\hline 
\end{tabular}

\caption{\label{tab:dw_oh}Representative comparison of domain-wall and one-hot encodings on four qubits. Note that, while convenient in this example, the convention of starting the domain-wall encoding at $0$ and one-hot at $1$ is not used elsewhere in the paper.}
\end{table}

\subsection{Binary Encoding}

While the main purpose of this work is to compare domain-wall and one-hot encodings, it is also worth comparing to direct binary encodings. Naively, these encodings appear that they should always be more efficient since to store a variable of size $m$ only requires $\lceil \log_2{m} \rceil $ binary variables as opposed to $m$ or $m-1$. However this does not tell the whole story, at currently available annealing hardware only has linear and quadratic interactions available. It is not necessarily true that the interactions between two binary variables can be written only in terms of these types of interactions, in fact except for special cases such as objective functions involving multiplication of variables \cite{chancellor2019,joseph2021} it is not going to be true. This can be rectified by engineering effective higher order interactions, \cite{Leib16a,chancellor17a} but each of these will require at least one auxilliary variable to engineer (although if this variable instantiated as a physical qubit it may have less stringent requirements than one used for computation \cite{chancellor17a}). A fair accounting of the binary variables needed for an encoding would also have to track these auxilliary variables as well. 

A lower bound of the number of auxilliary qubits for a given interaction can be obtained by a degree of freedom counting argument. For an arbitrary interaction of two variables both of size $m$, we require $m^2$ degrees of freedom, to independently assign energies to each configuration. On the other hand for $2\lceil \log_2{m} \rceil$ qubits, there will be $2\lceil \log_2{m} \rceil\times (1+2\lceil \log_2{m} \rceil)$ linear or quadratic degrees of freedom. The total number of higher than quadratic interactions needed will than be the difference between the number of degrees of freedom needed and the total number of linear or quadratic degrees of freedom available along with an additional $-1$ for the irrelevant energy offset degree of freedom,
\begin{equation}
n_{\mathrm{high}}(m)=m^2-
\lceil \log_2{m} \rceil (1+2\lceil \log_2{m} \rceil)-1.
\end{equation}
From the previous formula, we first see that for $m=2$ there are exactly the needed number of linear and quadratic degrees of freedom, which is a good check that this formula is sensible, since higher order terms are not possible in a system involving only two qubits. For $m=3$, there are more linear or quadratic terms than are necessary to encode all of the degrees of freedom, so auxilliary variables are not necessary for binary encoding. This is unsurprising, because the $m=3$ domain-wall encoding can be thought of as a special case of binary encoding, where the domain-wall interactions are used to eliminate one of the logical states and encode on the three remaining ones. 

\begin{figure}
    \includegraphics[width=0.5\textwidth]{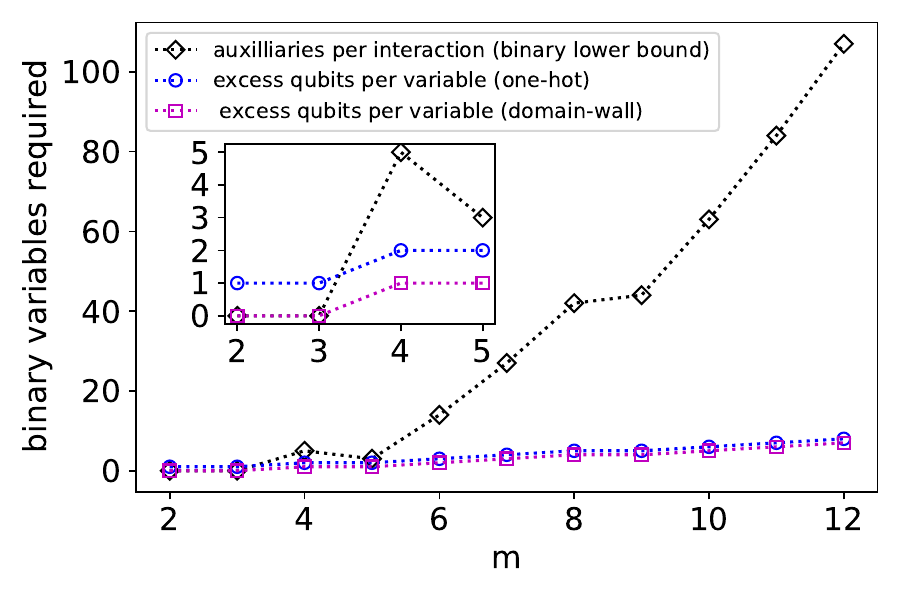}
        \caption[perform example]{Number of binary variables requires, per interaction in the binary case, and number of extra (when compared to binary) per higher variable in the domain-wall and one-hot cases, versus $m$. The inset is the same, but zoomed in to show detail at low values of $m$.} 
        \label{fig:excess_var_compare}
\end{figure}

However, for $m=4$, we find that there are $5$ fewer linear or quadratic degrees of freedom than are needed, and therefore $n_{\mathrm{high}}(4)=5$, indicating that each interaction between variables of size $m=4$ in a binary encoding will required at least $5$ auxilliary variable to engineer higher order interactions, and in practice may need more, for example the encoding of higher order interactions used in \cite{chancellor16a} requires $l$ auxilliary variable for an $l$ order interaction for any case where $l>3$. For simplicity we construct a lower bound assuming that each higher order interaction can be engineered using only one auxilliary. For a general problem with $q$ variables of size $m$ and $q_{\mathrm{int}}$ interactions between the variables, we therefore lower-bound the number of variables which are needed for the binary encoding as,
\begin{equation}
n_{\mathrm{bin}}(m,q,q_{\mathrm{int}})=q_{\mathrm{int}}n_{\mathrm{high}}(m)+q\lceil \log_2(m)\rceil. \label{eq:bin_bound}
\end{equation}
While we  examine the specific problems used in this study later, but for now let us consider the general question of when a binary encoding would be superior. Each higher-than-binary interaction will require at least one auxilliary binary variable, additionally, there have to be at least as many interactions as variables for the interaction graph of a problem to be connected. It is therefore reasonable to infer that if the number of auxilliary binary variables per interaction is greater than the number of excess binary variables to encode each higher variable when comparing domain-wall or one-hot to binary encoding than the binary encoding will not be favorable. As Fig.~\ref{fig:excess_var_compare} shows, the number of auxilliaries per interaction is always higher, indicating that at least in terms of the number of binary variables required to encode the problem, domain-wall and one-hot will be superior to direct binary encoding. Recall that this analysis only applies to arbitrary interactions and binary encoding may still be more efficient for specific kinds of interactions such as multiplication of numbers, however we are not aware of any such structure for the colouring problems studied here. This result is counter-intuitive, but somewhat unsurprising given that \cite{chancellor2019}, has previously argued that the domain-wall encoding is maximally efficient for encoding arbitrary interactions, assuming only linear and quadratic terms are directly accessible.

The present somewhat counter-intuitive picture where an encoding which is effectively unary is the most efficient way to represent arbitrary interactions depends entirely on the limitation in the types of coupling available (although we do note that this does meet the requirement of having a robust tensor product structure which is required for efficient quantum computing \cite{Blume-Kohout02a,chancellor2019}). This could change in the future if hardware with native higher order interactions and more exotic driver Hamiltonians (for example using the strategies of \cite{Leib16a,chancellor17a,Hen16a}) became available.  
\subsection{$k$-Coloring}One type of problem we use to contrast the performance of different QPU-encoding combinations, is maximum colouring problems. Formally these are  Max-$k$-Colorable Subgraph \cite{Hadfield17a} problems or equivalently Max-$k$-Cut problems \cite{Frieze97a,Khot07a}. However, since we do not consider any other types of coloring problems in the paper, we refer to these problems as $k$-coloring problems without fear of ambiguity. These  problems consist of finding ways to color a graph with $q$ nodes using $k$ colors such that nodes of the same color are in contact with each other in as few places as possible. For graphs which are colorable this reduces to finding a coloring of the graph. We use randomly generated Erd\"os-R\'enyi graphs \cite{erdos59a,erdos60a} where for each pair of nodes the presence or absence of an edge is independently and randomly decided with a given probability.

In the one-hot encoding scheme for the Max-$k$-Colorable Subgraph problem \cite{Hadfield17a},  we have $k=m$ decision variables $x_{i \alpha}$ for each node $i$ in the graph and every color $\alpha \in [m-1] $ which is $1$ if the node $i$ has the color $\alpha$.
The Hamiltonian for graph $G=([n-1], E)$ reads
\begin{equation} \label{eqn:graph_coloring_cost}
    H_\mathrm{DQM} = \sum_{\alpha=0}^{m-1} \sum_{(i, j) \in E} x_{i\alpha} x_{j\alpha} \,.
\end{equation}
Following the procedure in \cite{chancellor2019}, we consider two classes of $k$-coloring problems. The for the first we fix $k=3$ and vary $q$, we refer to these as maximum three coloring problems, or simply three coloring problems. For these problems we use graphs with an edge probability of $0.5$. 

The second class of coloring problems we consider are $k$-coloring problems where both $k$ and $q$ are varied, with $q=2 k$ and an edge probability of $0.75$, we refer to these as maximum $k$-coloring problems, or simply $k$-coloring problems. 

\begin{figure}
    \includegraphics[width=0.5\textwidth]{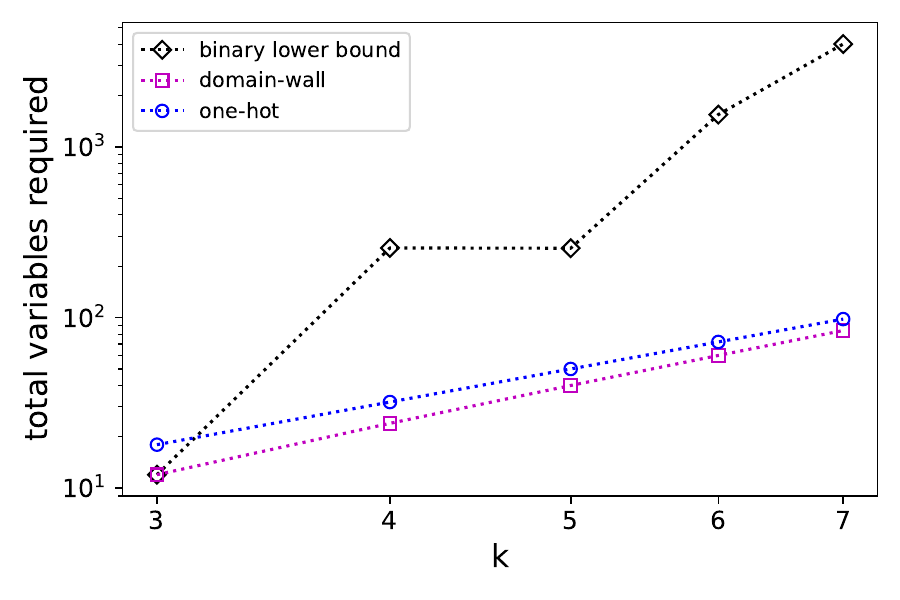}
        \caption[perform example]{The average number of binary variables required for domain-wall and binary encodings as well as a lower bound for binary encoding for $k$-colouring problems at sizes relevant to our experiments.} 
        \label{fig:binary_dw_oh_expt_sizes}
\end{figure}

For the maximum $k$-colouring problems, the the bound on the number of qubits needed from \ref{eq:bin_bound} will be $n_{\mathrm{bin}}(k,2k,0.75\times 4\times k^2)$. Throwing away non-leading order terms we find that asymptotically the number of variables required for the binary encoding scale as $k^4$ as opposed to $k^2$ for both domain-wall and one-hot. Fig.~\ref{fig:binary_dw_oh_expt_sizes} shows that even for $k=4$ binary encoding will require an impractically large number of variables. Recall that for size $k=3$ the domain-wall encoding is a special case of binary encoding.

\begin{figure}
    \includegraphics[width=0.5\textwidth]{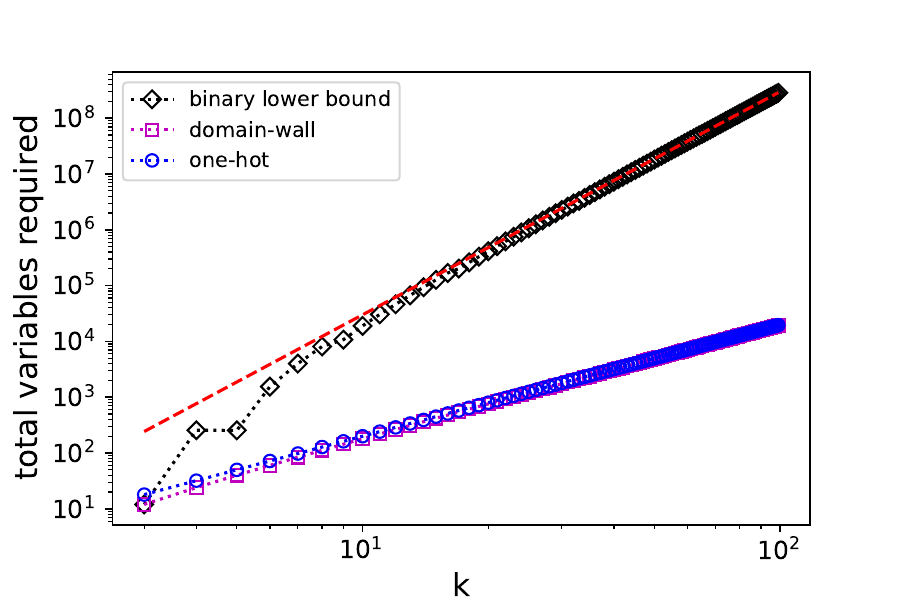}
        \caption[perform example]{The average number of binary variables required for domain-wall and binary encodings as well as a lower bound for binary encoding for $k$-colouring problems at sizes useful for showing approach to the asymptotic limit. The dashed line is the leading order term from the binary lower bound, $3k^4$.} 
        \label{fig:binary_dw_oh_scaling}
\end{figure}

We further extend the variable number calculation to higher $k$. As fig.~\ref{fig:binary_dw_oh_scaling} shows the binary encoding reaches a scaling which strongly resembles the final asymptotic scaling we expect and is far inferior to the domain-wall and even one-hot encoding.

\subsection{Flight-Gate Assignment \label{sub:flight_gate}}
The Flight-Gate Assignment problem was already investigated for QAOA~\cite{stollenwerk2020} and quantum annealing in the one-hot encoding scheme~\cite{stollenwerkFGA2019}.
We want to assign $n$ flights to $m$ gates using the 
decision variable
\begin{equation} \label{eqn:decision_variable}
	x_{i \alpha} = \begin{cases}
        1, &\text{if flight $i$ is assigned to gate $\alpha$}, \\
						0, &\text{otherwise},
					\end{cases}
\end{equation}
for $i \in [n-1]$ and $\alpha\in [m-1]$.
The cost function Hamiltonian calculates the total transit time of all passengers at the airport. It reads
\begin{equation} \label{eqn:cost}
    H_\textrm{DQM} = \sum_{i \alpha} \left( n_i^d t^d_\alpha + n_i^a t^a_\alpha \right) \, x_{i\alpha} + \sum_{ij\alpha\beta} n_{ij} t_{\alpha\beta} \, x_{i\alpha} \, x_{j\beta} \,
\end{equation}
where the various problem parameters are listed in Table~\ref{tab:fga_parameters}
\begin{table}[htpb]
    \centering
    \begin{tabularx}{\linewidth}{>{\hsize=0.2\hsize\linewidth=\hsize}X
        >{\hsize=0.8\hsize\linewidth=\hsize}X
    }
        \toprule
        Symbol              & Meaning \\
        \midrule
        $n\dep_i$    		& \# of passengers departing with flight $i$ \\
        $n\arr_i$      		& \# of passengers arriving with flight $i$ \\
        $n_{ij}$     		& \# of transfer passengers between flights $i$ and $j$ \\
         $t\iin_i$ 			& Arrival time of flight $i$ \\
        $t\out_i$ 			& Departure time of flight $i$ \\
        $t\arr_\alpha$ 		& Transfer time from gate $\alpha$ to baggage claim \\
        $t\dep_\alpha$ 		& Transfer time from check-in to gate $\alpha$ \\
        $t_{\alpha\beta}$ 	& Transfer time from gate $\alpha$ to gate $\beta$ \\
        $t\buf$			 	& Buffer time between two flights at the same gate \\
        \bottomrule
    \end{tabularx}
    \caption{Flight-Gate Assignment problem parameter}\label{tab:fga_parameters}
\end{table}

There are two hard constraints in the problem. 
First, a flight should be assigned to exactly one gate.
This is represented by the Equation~\eqref{eq:constraint}.
Second, flights with temporal overlap are not allowed to be assigned to the same gate.
\begin{equation}
    \forall \alpha,  \forall (i, j) \in E: \qquad x_{i, \alpha} \cdot x_{j, \alpha} = 0 \, ,
\end{equation}
where
\begin{equation}\label{eqn:forbidden_flight_pairs}
    E = \left\{(i, j) \; | \; (t^{in}_i - t^{out}_j < t^\text{buf} ) \wedge  (t^{in}_j - t^{out}_i < t^\text{buf} )  \right\} \, 
\end{equation}
is the set of forbidden flight pairs.
This second constraint is enforced by adding the following 
\begin{equation} \label{eq:fga_constraint}
    H_\text{temp} = \mu \sum_\alpha \sum_{(i, j) \in E} x_{i\alpha} x_{j\alpha}  \; ,
\end{equation}
to the total Hamiltonians~\eqref{eq:Hoh} and~\eqref{eq:Hdw}, respectively.
Again the penalty weight $\mu$ must be sufficiently large to ensure the constraint satisfaction in the ground state (see~\cite{stollenwerkFGA2019} for details).
Note that both constraints are equivalent to the proper coloring of a graph with edged $E$. 
This relation to graph coloring is extensively discussed e.g.\ in~\cite{stollenwerk2020}.

Due to precision problems of the D-Wave quantum annealer, in~\cite{stollenwerkFGA2019} it was found, that the largest instances with non-vanishing success probability where 29 instance with $n=7$ flights and $m=2$.
We used these instances for our study. Given that no variable in the flight gate assignment problems studied here is larger than $m=3$, all domain-wall encodings used for these problems are also special cases of binary encoding. 
 
 \section{Experimental Setup and Performance Measures}
In this section, we discuss the experimental setup and measures for comparing the performance of domain-wall and one-hot encoding.
\subsection{Chain and constraint strengths}
To experimentally study these problems we need to decide upon a method to choose the strength of both
embedding chains and the constraints used to restrict the number of domain walls to one or enforce the
one-hot constraint (e.g. $\lambda$, $\mu$ and $\kappa$ in Equations~\eqref{eq:Hoh},~\eqref{eq:Hdw} and~\eqref{eq:fga_constraint}). Since the goal of this paper is to compare rather than develop an absolute benchmark, it is not crucial that these choices be completely optimal, but for these comparisons to be relevant to real calculations we should still ensure that we are operating in a regime where the behavior is likely to be similar to the optimal choices.

\begin{figure}
    \includegraphics[width=0.5\textwidth]{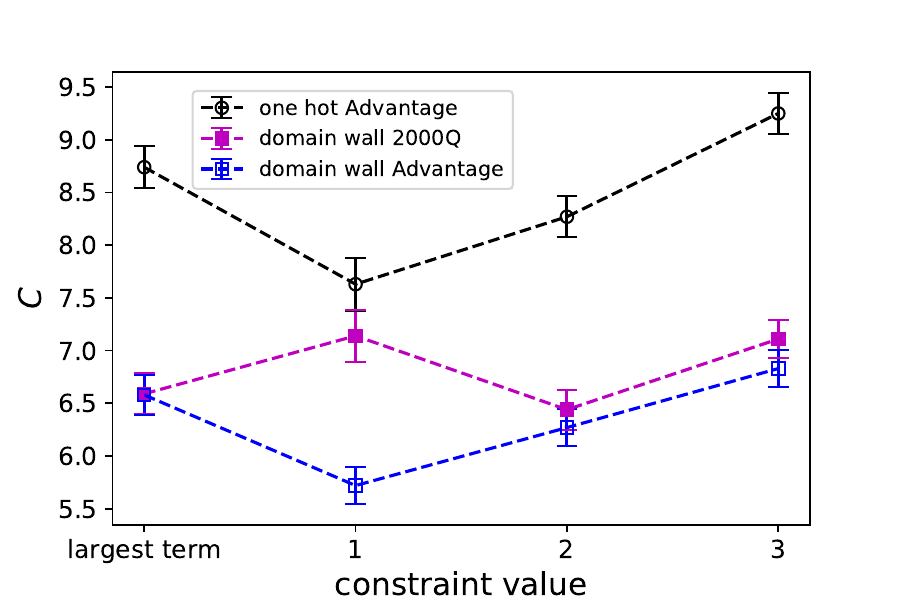}
        \caption[perform example]{The cost function C, which is equal to the total number of places which the same color touch for the three coloring problem with $15$ nodes, averaged over $100$ instances versus the constraint strength.} 
        \label{fig:constraint_performance}
\end{figure}

For the chain strength we use the ``uniform torque compensation'' \cite{uniformTorque} feature which is available through the
D-Wave ocean software repository \cite{ocean}. On the other hand no such
feature exists for finding the strength of the constraints. 
However a practical approach is to 
choose the strength of the constraint parameters $\lambda$, $\kappa$ and $\mu$
equal to the magnitude of the largest single field or coupler in the problem definition before embedding.
Both of these choices have the advantage of being ``automatic'' the sense of adjusting based on the structure of the problem at hand. Since the goal of this paper is to make a comparison of performance on the same footing, rather than to test the optimal performance of the device, we do not need to show that our parameter choices are the best possible, but that they perform reasonably well.  To verify that these choices are sensible, we compare different constraint choices for a size $15$ three coloring problem in Figure~\ref{fig:constraint_performance}. We find that our strategy for choosing the constraint strength is roughly optimal for 2000Q, it is slightly suboptimal for Advantage. While this performs well enough for the purposes of the analysis we do here, it is worth keeping in mind that there is potential room for further gains within our Advantage data by finding a better performing parameter setting heuristic.

\subsection{Experimental procedures}
All experiments reported here were performed in the autumn of 2020. They were performed using the default anneal time of $5 \mu$s and $100$ anneals were performed in each run using a single embedding (we found that the performance variation between different embeddings was negligible). All embeddings were performed using the minor-miner embedding software provided by D-Wave systems Inc.~\cite{minorminer}. If an embedding failed, we attempt it twice more to verify that this failure was not an anomaly. We found that for all problem classes (for example particular size of graph for three coloring or number of colors for k-coloring), for any given QPU/embedding combination either all problem embeddings failed or they all succeeded. 
Except when explicitly stated otherwise in the appendix, broken chain decoding was performed using the majority vote decoding tool included in the software package. Spin reversal transforms (sometimes referred to as gauge averaging) were not used. Experimental data are available in a public repository \cite{domain_performance_data}.

Numerical analysis and plotting were performed using Matlab and the python programming language \cite{van2003python}, in particular, heavy use was made of the numpy \cite{numpy,oliphant2006guide} and matplotlib \cite{hunter2007matplotlib} packages, as well as jupyter notebooks \cite{jupyter,perez2007ipython}.

\subsection{Hypothesis testing}\label{sec:hypothesis_testing}
One method to compare between pairs of QPU-encoding combinations is to run them on the same problems and see how many times each can outperform the others, ignoring cases where they can each find equally good solutions. The immediate question then becomes how statistically significant a given sample is. 
In other words, how likely are we to see a result which is at least as favorable by random chance? To quantify this significance, we perform hypothesis testing. 

To this end, we partition the solutions where one QPU-encoding combination outperforms the other into the two classes.
First, let the number of cases where the combination we expect to do better (the ``expected winner'') has performed better than the combination it is being compared to $n_b$.
And second, let the number of cases where the ``expected winner'' performs worse $n_w$. 
This allows us to calculate the statistical significance,
\begin{equation} \label{eq:signifcance}
    p=\frac{1}{2^{n_b+n_w}}\sum^{n_b+n_w}_{k=n_w} \binom{n_b+n_w}{k} \, . 
\end{equation}
This is effectively the probability that the expected winner could perform better at least as many times if the better performing QPU-encoding combination were chosen at random with 50\% probability (our null hypothesis). Effectively, it is the probability that the result (or an even more favorable one) could happen by chance if both performed equally well.

By convention, $p<0.05$ is considered to be a statistically significant result rejecting the null hypothesis \cite{Lehmann93a} and therefore confirming that the expected winner does indeed perform better, by symmetry, $p>0.95$ is also a statistically significant result, but rejecting an alternate hypothesis that the expected winner was chosen correctly, and therefore showing that we chose the expected winner incorrectly. A value of $0.05<p<0.95$ is not statistically significant and indicates that an insufficient number of samples have been taken to draw any conclusions based on our hypothesis testing strategy. 
\subsection{Performance Measures\label{sec:measures}}

To understand the effect of encoding and QPU choice on the ability of the device to solve the problem, we choose to analyze four performance metrics. The first is the fraction of raw solutions which do not contain any broken embedding chains, $R_\mathrm{chain}$. This measure allows us to get a sense of how faithfully the device is able to represent the problem, since solutions with broken chains no longer correspond to valid solutions.

The second quantity we examine is the rate of solutions where, after majority vote decoding is performed on any broken chains, the solutions satisfy all of the one-hot or domain-wall constraints.We call this rate $R_\mathrm{enc}$. 

We next define the cost function $C$ which is the quantity which we are attempting to minimize subject to our constraints. By convention, if an annealing run has not returned any solutions where all the constraints are satisfied, then we define the effective value of $C$ to be infinite, since no valid solution is returned, therefore as we have defined it, the average cost function is infinite even if a single problem does not yield a valid solution.

Finally we define the success probability $P$, which is defined as the fraction of problems for which the optimal value of $C$ was found. We only report this measure for problem sizes which are small enough that the optimal solution can be definitively found using exhaustive search. Recall that we only perform $100$ anneals per run, so it is likely that higher success probabilities could be found by increased sampling, potentially also using spin reversal transforms, or even more advanced tricks like reverse annealing \cite{Perdomo-Ortiz11,chancellor17b,Venturelli19a}, pausing \cite{Marshall19a}, sample persistence \cite{Karimi17a}, anneal offsets \cite{Yarkoni19a}, or extended coupling range \cite{virtGraph}. The goal of this study is a relative comparison between QPUs and encoding methods rather than to benchmark the best possible performance which can be attained using these devices, and for this reason we have elected to keep our experiments simple at the expense of cutting-edge performance.
 
\section{Results}

In this section, we present our results by showing the four performance measures for the flight gate assignment problem, the three-colouring problem and $k$-colouring problem.
Also, we show our results on the hypothesis testing for the three-colouring and $k$-colouring problem.

\subsection{Flight Gate Assignment}

\begin{figure*}[htp!]
        \centering
        \begin{subfigure}[b]{0.475\textwidth}
            \centering
            \includegraphics[width=\textwidth]{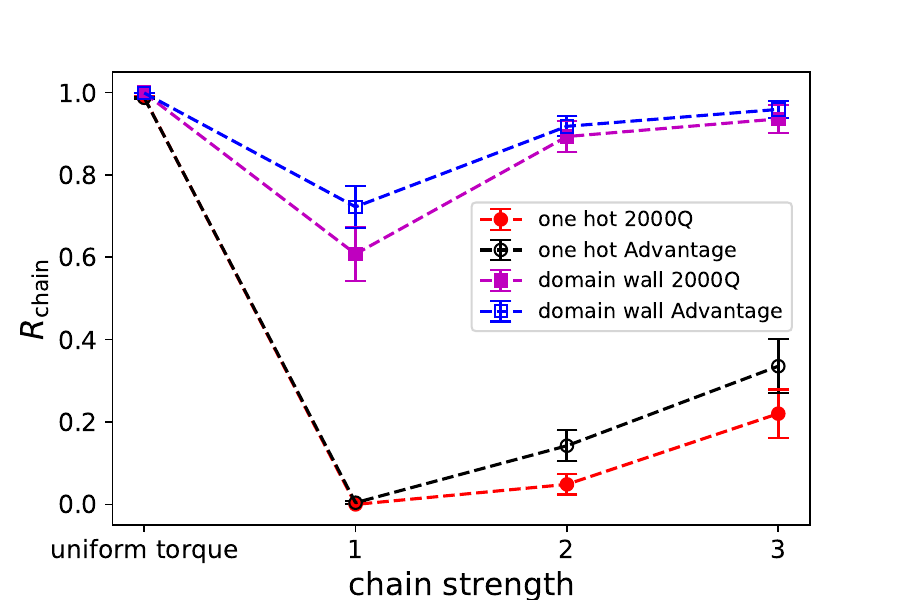}
            \caption[Rate of unbroken chains in physical solutions]{Rate of unbroken chains in physical solutions}
            \label{fig:flight_chain}
        \end{subfigure}
        \hfill
        \begin{subfigure}[b]{0.475\textwidth}  
            \centering 
            \includegraphics[width=\textwidth]{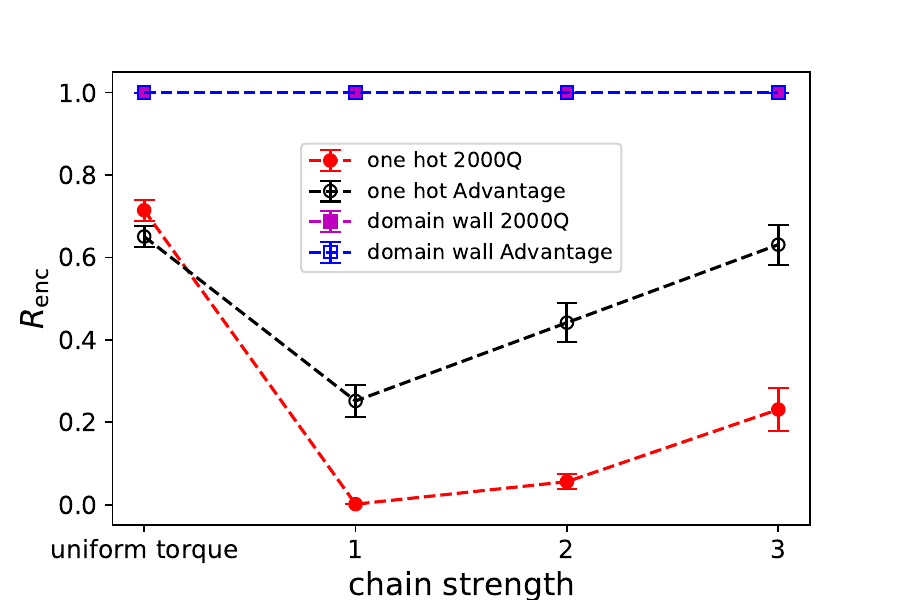}
            \caption[Rate of correctly encoded logical solutions]{Rate of correctly encoded logical solutions}   
            \label{fig:3flight_enc}
        \end{subfigure}
        \vskip\baselineskip
        \begin{subfigure}[b]{0.475\textwidth}   
            \centering 
            \includegraphics[width=\textwidth]{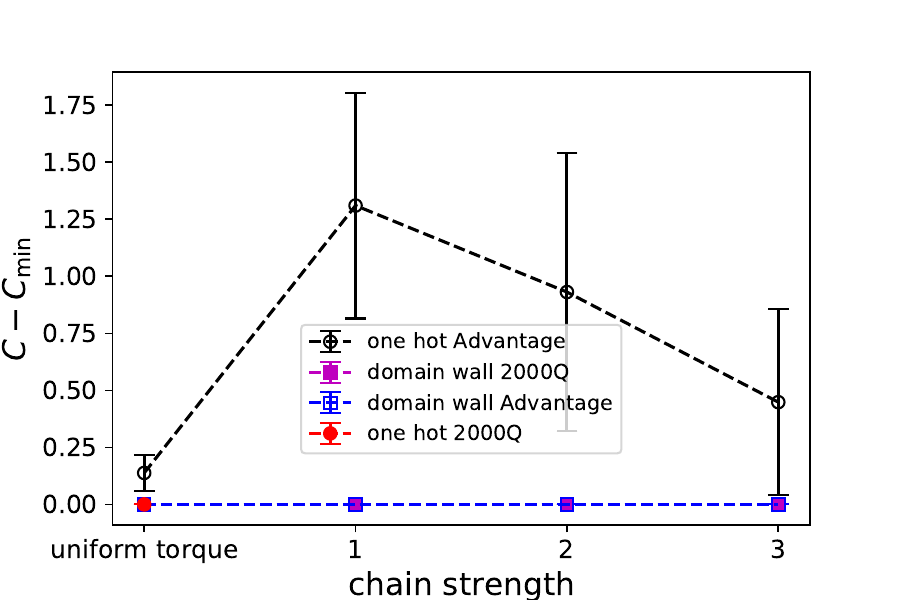}
            \caption[Cost function of the problem from logical solutions]{Cost function of the problem from logical solutions}    
            \label{fig:flight_cost}
        \end{subfigure}
        \hfill
        \begin{subfigure}[b]{0.475\textwidth}   
            \centering 
            \includegraphics[width=\textwidth]{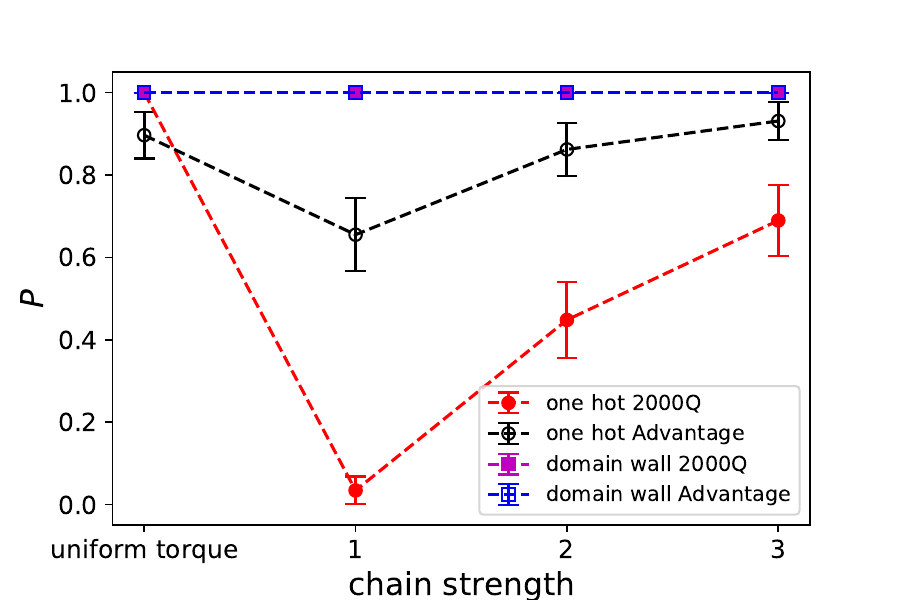}
            \caption[]{Success probability}    
            \label{fig:flight_success}
        \end{subfigure}
        \caption[3-coloring Problem]
        {Mean of the four performance measures against the chain strength for the $29$ instances of the flight gate assignment problem for all four QPU-encoding combinations. 
        Note that the cost function is normalized by subtracting the optimal value $C_{\mathrm{min}}$ for more meaningful comparison across problems. 
        In part~(\subref{fig:flight_cost}) the one-hot 2000Q point only appears for uniform torque compensation since the data for all other chain strengths contained at least one problem where none of the $100$ reads produced a solution which satisfied all one-hot constraints. 
    All bars in this plot are standard error.}
        \label{fig:fg}
\end{figure*}
As discussed in section~\ref{sub:flight_gate}, all studied flight gate assignment instances have $m=2$. 
Since the encoding of a discrete variable of size two into a domain-wall encoding reduces to a direct binary encoding, it is not mathematically possible for the domain-wall constraint to be violated in these cases. 
On the other hand, we find that the one-hot constraint is only satisfied in $71\%$ and $65\%$ of solutions on average for 2000Q and Advantage respectively. 
We do find some chain breaks in all cases, but they are so rare it is not possible to reliably differentiate which encoding performs better based on our data, although we have observed that the domain-wall encoding seems to perform slightly better. 
We observe that the problem is solved after $100$ reads in all cases except for three using the one-hot encoding on Advantage. 

To study what effect chain breaks would have on larger systems, we set embedding chain strengths to intentionally suboptimal values as opposed to using the uniform torque compensation tool, which performs well in all cases. 
The results are shown in Figure~\ref{fig:fg}. 
We find that the domain-wall encoding performs equally or better than one-hot against all metrics, regardless of the QPU used. 
What is particularly striking is that the domain-wall encoding was able to solve all instances at all chain strengths for both processors, while the one-hot encoding was only able to achieve this for the uniform torque compensation scheme.

\subsection{Three-Coloring}

\begin{figure*}[htp!]
        \centering
        \begin{subfigure}[b]{0.475\textwidth}
            \centering
            \includegraphics[width=\textwidth]{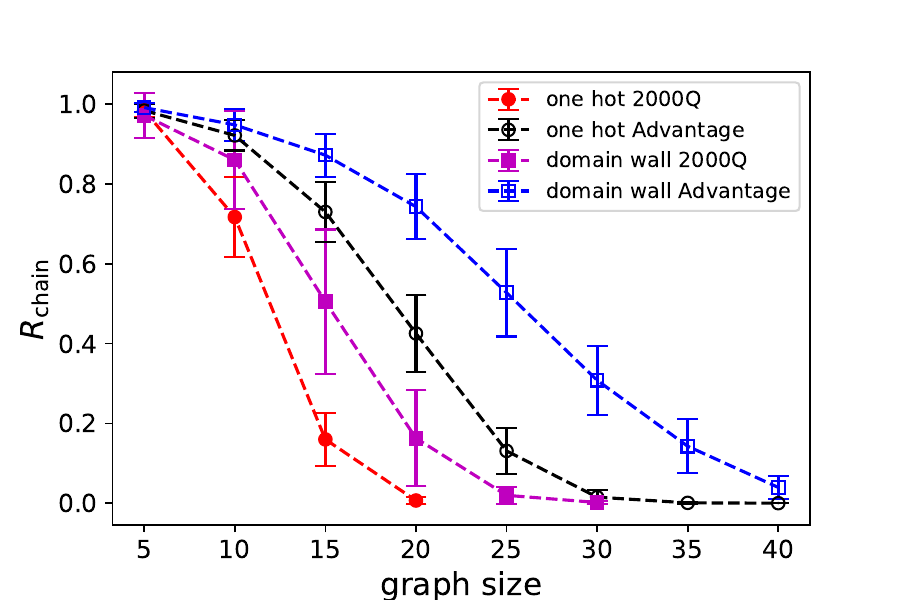}
            \caption[Rate of unbroken chains in physical solutions]{Rate of unbroken chains in physical solutions}
            \label{fig:3col_chain}
        \end{subfigure}
        \hfill
        \begin{subfigure}[b]{0.475\textwidth}  
            \centering 
            \includegraphics[width=\textwidth]{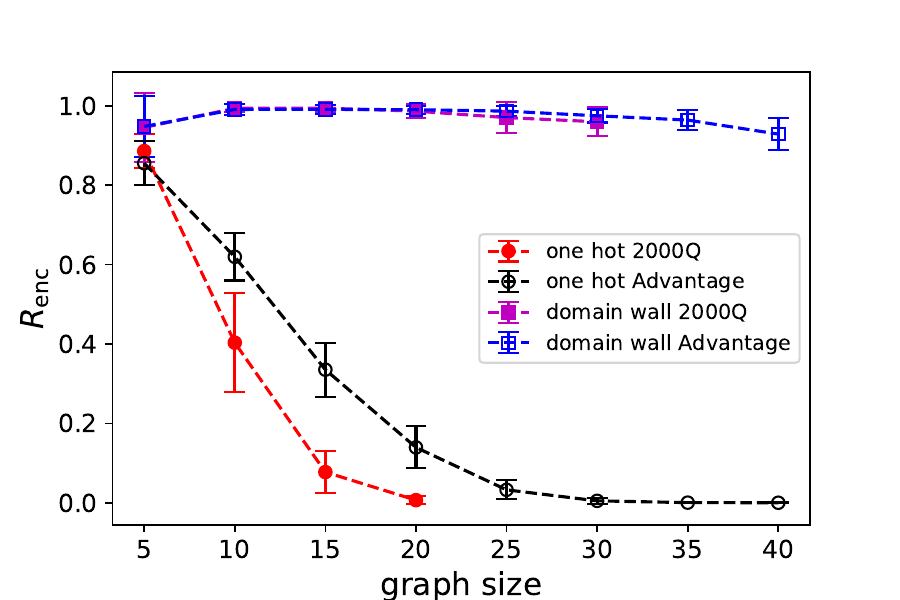}
            \caption[Rate of correctly encoded logical solutions]{Rate of correctly encoded logical solutions}   
            \label{fig:3col_enc}
        \end{subfigure}
        \vskip\baselineskip
        \begin{subfigure}[b]{0.475\textwidth}   
            \centering 
            \includegraphics[width=\textwidth]{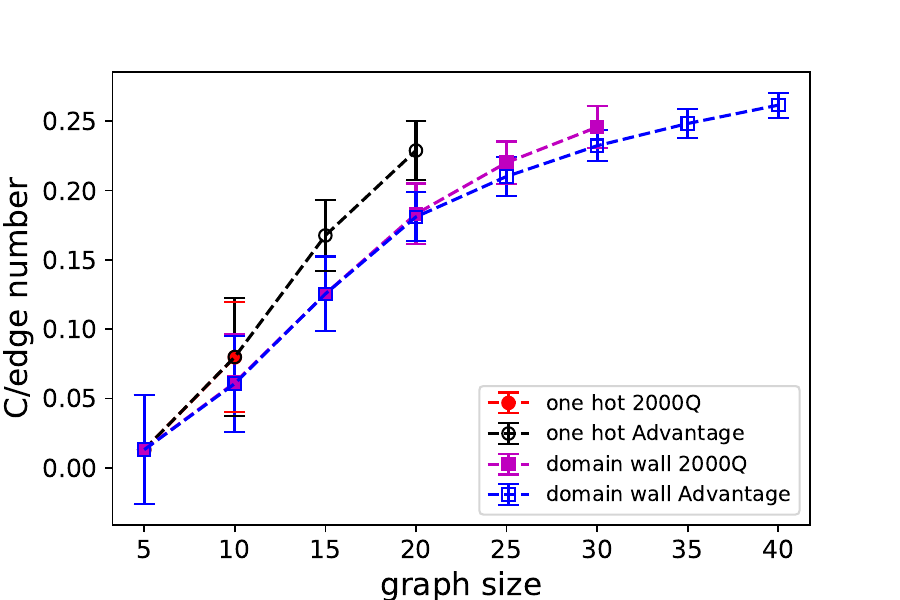}
            \caption[Cost function of the problem from logical solutions]{Cost function of the problem from logical solutions}    
            \label{fig:3col_cost}
        \end{subfigure}
        \hfill
        \begin{subfigure}[b]{0.475\textwidth}   
            \centering 
            \includegraphics[width=\textwidth]{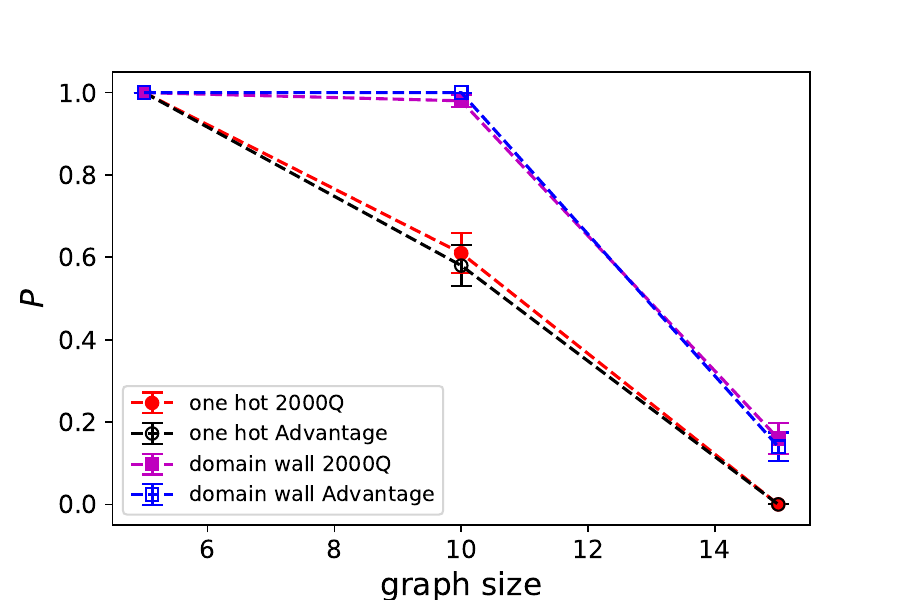}
            \caption[]{Success probability}    
            \label{fig:3col_success}
        \end{subfigure}
        \caption{Mean of the four performance measures against the problem size for the three coloring problem  for all four QPU-encoding combinations. 
        The cost function in part~(\subref{fig:3col_cost}) has been normalized by edge number to allow a more direct comparisons at different sizes. 
        The bars for parts~(\subref{fig:3col_chain}),~(\subref{fig:3col_enc}) and~(\subref{fig:3col_cost}) are the standard deviation of the distribution, rather than standard error. 
        Since all data points are based on $100$ samples, the standard error is ten times smaller than what is depicted by these bars. 
    Bars for (\subref{fig:3col_success}) are standard error.}
        \label{fig:3col}
\end{figure*}

For further demonstration of the effect of the different encodings, we consider $100$ random
instances of three color problems as studied in~\cite{chancellor2019}. For these problems each
variable $d_i$ has three possible values, corresponding to each color, since the domain-wall encoding
consists of two qubits in this case, it is mathematically possible for the single domain-wall
constraint to not be satisfied. For each of these instances we
run both domain-wall and one-hot encodings each on the Advantage and 2000Q QPU. 

Figure~\ref{fig:3col} shows the four performance measures as discussed in section~\ref{sec:measures}.
For the same QPU the domain-wall encoding performs consistently better than one-hot.
Also, for the same encoding the Advantage QPU performs consistently better than the 2000Q QPU.
When comparing the domain-wall encoding on the 2000Q QPU with one-hot encoding on the Advantage QPU, we can show that the average solution quality appears to be either comparable or to favor the domain-wall on the less advanced processor. 
On the other hand, even using the one-hot encoding, the Advantage processor has fewer chain breaks. 
In essence, using a more connected QPU and using domain-wall as opposed to one-hot seem to reduce chain breaks and furthermore, the higher connectivity seems to be the more decisive factor. 

However, it seems to be the case that (at least for majority vote decoding), broken chains in the domain-wall encoding are more likely to be decoded to correct solutions. For the cost function as well, there is a visible difference between the performance of both the one-hot and domain-wall encoding. Finally, we observe that, while the domain-wall encoding makes a large difference in the probability of finding the optimal solution within $100$ reads, the difference between the two types of QPUs is within error bars for all sizes we test. Aside from the fact that larger problems can be encoded on the Advantage QPU, there is not a significant difference between the performance of the two QPUs for the same encoding for graph sizes less than about $25$, however as we demonstrate later, the performance differences can be better understood using analysis based on hypothesis testing. 

\subsection{Hypothesis Testing for Three-Coloring}
\begin{table*}[t]

\begin{tabular}{|c|c|c|c|c|c|c|c|c|c|c|c|c|} 
 \hline& \multicolumn{2}{c|}{Adv. dw/oh} & \multicolumn{2}{c|}{2000Q dw/oh} & \multicolumn{2}{c|}{dw Adv./2000Q} & \multicolumn{2}{c|}{oh Adv./2000Q} & \multicolumn{2}{c|}{(dw, Adv.)/(oh, 2000Q)} & \multicolumn{2}{c|}{(dw, 2000Q)/(oh, Adv.)}\tabularnewline 
 \hline 
 \hline 
5 node (b,w) & \hspace{0.2 cm} 0 \hspace{0.2 cm} & 0 &  \hspace{0.2 cm} 0 \hspace{0.2 cm}     & 0 &  \hspace{0.2 cm} 0 \hspace{0.2 cm}  & 0 &  \hspace{0.2 cm} 0 \hspace{0.2 cm}  & 0 &  \hspace{0.2 cm} 0 \hspace{0.2 cm}  & 0 &  \hspace{0.2 cm} 0 \hspace{0.2 cm}  & 0 \tabularnewline 
 \hline 
5 node  p & \multicolumn{2}{c|}{ } & \multicolumn{2}{c|}{ } & \multicolumn{2}{c|}{ } & \multicolumn{2}{c|}{ } & \multicolumn{2}{c|}{ } & \multicolumn{2}{c|}{ }  \tabularnewline 
 \hline 
10 node (b,w) & \hspace{0.2 cm} 42 \hspace{0.2 cm} & 0 &  \hspace{0.2 cm} 37 \hspace{0.2 cm}     & 0 &  \hspace{0.2 cm} 2 \hspace{0.2 cm}  & 0 &  \hspace{0.2 cm} 19 \hspace{0.2 cm}  & 21 &  \hspace{0.2 cm} 39 \hspace{0.2 cm}  & 0 &  \hspace{0.2 cm} 40 \hspace{0.2 cm}  & 0 \tabularnewline 
 \hline 
10 node  p & \multicolumn{2}{c|}{$2.27\times 10^{-13}$\cellcolor{green!50}} & \multicolumn{2}{c|}{$7.28\times 10^{-12}$\cellcolor{green!50}} & \multicolumn{2}{c|}{$2.50\times 10^{-1}$\cellcolor{yellow!50}} & \multicolumn{2}{c|}{$6.82\times 10^{-1}$\cellcolor{yellow!50}} & \multicolumn{2}{c|}{$1.82\times 10^{-12}$\cellcolor{green!50}} & \multicolumn{2}{c|}{$9.09\times 10^{-13}$\cellcolor{green!50}}  \tabularnewline 
 \hline 
15 node (b,w) & \hspace{0.2 cm} 85 \hspace{0.2 cm} & 2 &  \hspace{0.2 cm} 95 \hspace{0.2 cm}     & 3 &  \hspace{0.2 cm} 32 \hspace{0.2 cm}  & 34 &  \hspace{0.2 cm} 70 \hspace{0.2 cm}  & 22 &  \hspace{0.2 cm} 94 \hspace{0.2 cm}  & 1 &  \hspace{0.2 cm} 91 \hspace{0.2 cm}  & 2 \tabularnewline 
 \hline 
15 node  p & \multicolumn{2}{c|}{$2.47\times 10^{-23}$\cellcolor{green!50}} & \multicolumn{2}{c|}{$4.95\times 10^{-25}$\cellcolor{green!50}} & \multicolumn{2}{c|}{$6.44\times 10^{-1}$\cellcolor{yellow!50}} & \multicolumn{2}{c|}{$2.67\times 10^{-7}$\cellcolor{green!50}} & \multicolumn{2}{c|}{$2.42\times 10^{-27}$\cellcolor{green!50}} & \multicolumn{2}{c|}{$4.41\times 10^{-25}$\cellcolor{green!50}}  \tabularnewline 
 \hline 
20 node (b,w) & \hspace{0.2 cm} 99 \hspace{0.2 cm} & 0 &  \hspace{0.2 cm} 100 \hspace{0.2 cm}     & 0 &  \hspace{0.2 cm} 43 \hspace{0.2 cm}  & 41 &  \hspace{0.2 cm} 94 \hspace{0.2 cm}  & 3 &  \hspace{0.2 cm} 100 \hspace{0.2 cm}  & 0 &  \hspace{0.2 cm} 93 \hspace{0.2 cm}  & 2 \tabularnewline 
 \hline 
20 node  p & \multicolumn{2}{c|}{$1.58\times 10^{-30}$\cellcolor{green!50}} & \multicolumn{2}{c|}{$7.89\times 10^{-31}$\cellcolor{green!50}} & \multicolumn{2}{c|}{$4.57\times 10^{-1}$\cellcolor{yellow!50}} & \multicolumn{2}{c|}{$9.60\times 10^{-25}$\cellcolor{green!50}} & \multicolumn{2}{c|}{$7.89\times 10^{-31}$\cellcolor{green!50}} & \multicolumn{2}{c|}{$1.15\times 10^{-25}$\cellcolor{green!50}}  \tabularnewline 
 \hline 
25 node (b,w) & \hspace{0.2 cm} 100 \hspace{0.2 cm} & 0 &  \hspace{0.2 cm}   &  FAIL &  \hspace{0.2 cm} 66 \hspace{0.2 cm}  & 20 &  \hspace{0.2 cm}   &  FAIL &  \hspace{0.2 cm}   &  FAIL &  \hspace{0.2 cm} 98 \hspace{0.2 cm}  & 2 \tabularnewline 
 \hline 
25 node  p & \multicolumn{2}{c|}{$7.89\times 10^{-31}$\cellcolor{green!50}} & \multicolumn{2}{c|}{ } & \multicolumn{2}{c|}{$3.33\times 10^{-7}$\cellcolor{green!50}} & \multicolumn{2}{c|}{ } & \multicolumn{2}{c|}{ } & \multicolumn{2}{c|}{$3.98\times 10^{-27}$\cellcolor{green!50}}  \tabularnewline 
 \hline 
30 node (b,w) & \hspace{0.2 cm} 100 \hspace{0.2 cm} & 0 &  \hspace{0.2 cm}   &  FAIL &  \hspace{0.2 cm} 72 \hspace{0.2 cm}  & 20 &  \hspace{0.2 cm}   &  FAIL &  \hspace{0.2 cm}   &  FAIL &  \hspace{0.2 cm} 97 \hspace{0.2 cm}  & 2 \tabularnewline 
 \hline 
30 node  p & \multicolumn{2}{c|}{$7.89\times 10^{-31}$\cellcolor{green!50}} & \multicolumn{2}{c|}{ } & \multicolumn{2}{c|}{$2.30\times 10^{-8}$\cellcolor{green!50}} & \multicolumn{2}{c|}{ } & \multicolumn{2}{c|}{ } & \multicolumn{2}{c|}{$7.81\times 10^{-27}$\cellcolor{green!50}}  \tabularnewline 
 \hline 
35 node (b,w) & \hspace{0.2 cm} 100 \hspace{0.2 cm} & 0 &  FAIL &  FAIL &  \hspace{0.2 cm}   &  FAIL &  \hspace{0.2 cm}   &  FAIL &  \hspace{0.2 cm}   &  FAIL &  FAIL &   \tabularnewline 
 \hline 
35 node  p & \multicolumn{2}{c|}{$7.89\times 10^{-31}$\cellcolor{green!50}} & \multicolumn{2}{c|}{ } & \multicolumn{2}{c|}{ } & \multicolumn{2}{c|}{ } & \multicolumn{2}{c|}{ } & \multicolumn{2}{c|}{ }  \tabularnewline 
 \hline 
40 node(b,w) & \hspace{0.2 cm} 100 \hspace{0.2 cm} & 0 &  FAIL &  FAIL &  \hspace{0.2 cm}   &  FAIL &  \hspace{0.2 cm}   &  FAIL &  \hspace{0.2 cm}   &  FAIL &  FAIL &   \tabularnewline 
 \hline 
40 node p & \multicolumn{2}{c|}{$7.89\times 10^{-31}$\cellcolor{green!50}} & \multicolumn{2}{c|}{ } & \multicolumn{2}{c|}{ } & \multicolumn{2}{c|}{ } & \multicolumn{2}{c|}{ } & \multicolumn{2}{c|}{ }  \tabularnewline 
 \hline 
\end{tabular} 
 
\caption{\label{tab:hypo_three}Hypothesis testing results for all six possible comparisons of QPU-encoding combinations for three color problems of different sizes. 
For each comparison the expected winner is listed first. 
For each size the count of cases where the expected winner (written first at the top of the column) performs better $n_b$ (left) and worse $n_w$ (right) are listed. 
Below is listed the value of $p$ as calculated by Equation~\eqref{eq:signifcance}. 
In cases where either both combinations perform the same on all problems, or one or both fail to embed, statistical significance cannot be calculated. 
In case where the expected winner failed to embed, we write `FAIL' in the left column, and likewise if the embedding fails for the QPU-encoding combination described by the right column. 
These comparisons are performed for the single best solution found out of all $100$ samples, using majority vote decoding for broken chains. 
If none of the samples decode to valid solutions, than the cost function is treated as being ``infinite'' and any finite value is considered to be better. 
Color coding used as a guide to the eye, green indicates a statistically significant rejection of the null hypothesis, while yellow indicates a result which is not statistically significant.}
\end{table*}

Since it is difficult to visually distinguish the average cost functions, we apply a different technique to compare performance of solutions against the cost function. We examine how many cases each processor-encoding combination does better or worse than any of the others. We further perform hypothesis testing to see which of the differences are statistically significant as described in the Section~\ref{sec:hypothesis_testing}. 
The results are shown in Table~\ref{tab:hypo_three}. 
We can see that except for at the smallest size (5 nodes), there are differences in how the processor-encoding combinations perform. We find that except for at this smallest size, the domain-wall encoding always performs better than the one-hot, even when comparing the domain-wall encoding on a 2000Q to one-hot encoding on an Advantage. We further find that all statistically significant results point toward Advantage performing better than 2000Q, but at smaller sizes the differences are not statistically significant.

A particularly striking result here is that, at least up to the size where the problems can no longer be embedded on the 2000Q, using a domain-wall, rather than one-hot encoding makes a bigger difference to solution quality than using the more advanced processor. This underscores the importance of encoding methods to obtaining high quality solutions over just waiting for hardware improvements. An astute reader may question whether this result is simply because the majority vote decoding seems to perform better on domain-wall encoded problems than one-hot (as can be seen by comparing Figure~\ref{fig:3col_chain} to Figure~\ref{fig:3col_enc}). To answer this question, we consider an alternate way of processing the data, in which solutions with broken chains are discarded rather than decoded by majority vote. We find that this approach does not affect the qualitative result that the domain-wall encoding on a 2000Q performs better in a statistically significant way. For completeness, these results are shown in Table~\ref{tab:hypo_three_noDec} in the appendix. While the domain-wall encoding does always make a bigger performance difference in cases where the problem can be embedded on both QPUs, it is possible to embed larger problems on the Advantage QPU, even with the domain-wall encoding, $30$ nodes is the largest size we are able to embed on an 2000Q, whereas Advantage can embed a problem of at least size $40$.

\begin{figure*}[htp!]
        \centering
        \begin{subfigure}[b]{0.475\textwidth}
            \centering
            \includegraphics[width=\textwidth]{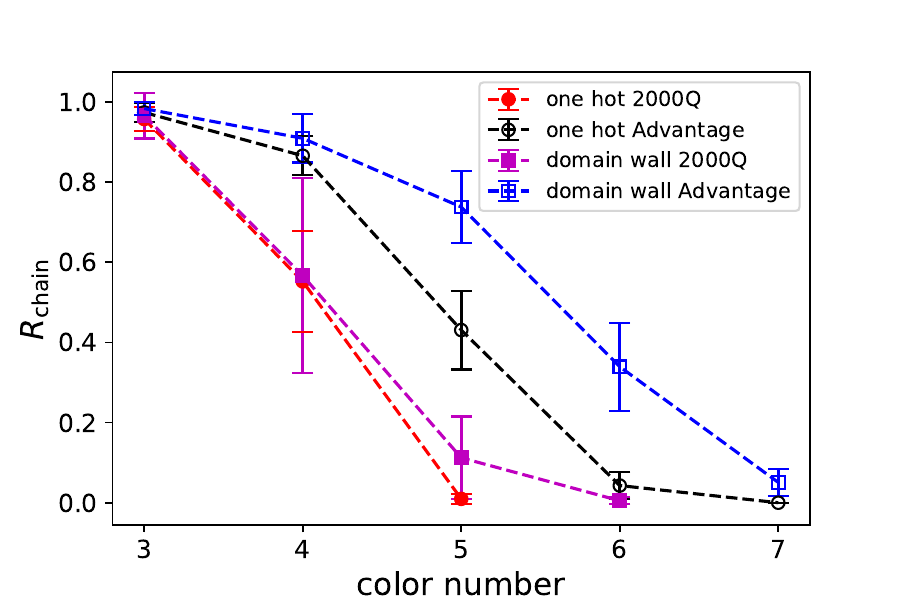}
            \caption[Rate of unbroken chains in physical solutions]{Rate of unbroken chains in physical solutions}
            \label{fig:kcol_chain}
        \end{subfigure}
        \hfill
        \begin{subfigure}[b]{0.475\textwidth}  
            \centering 
            \includegraphics[width=\textwidth]{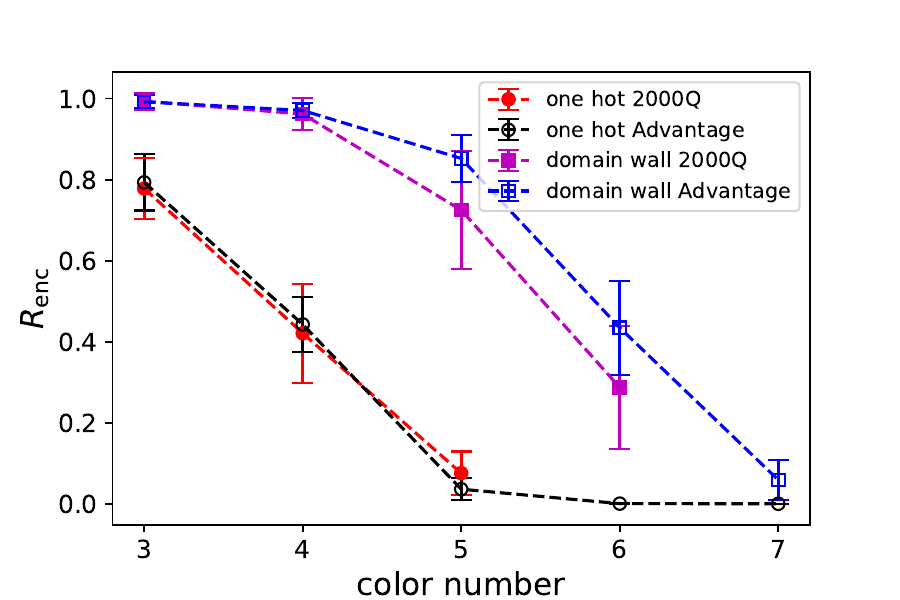}
            \caption[Rate of correctly encoded logical solutions]{Rate of correctly encoded logical solutions}   
            \label{fig:kcol_enc}
        \end{subfigure}
        \vskip\baselineskip
        \begin{subfigure}[b]{0.475\textwidth}   
            \centering 
            \includegraphics[width=\textwidth]{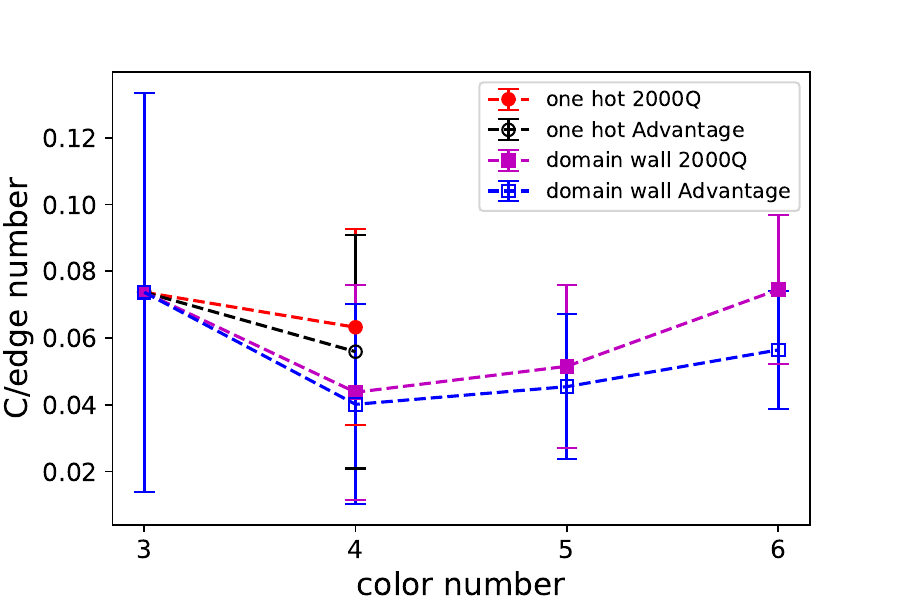}
            \caption[Cost function of the problem from logical solutions]{Cost function of the problem from logical solutions}    
            \label{fig:kcol_cost}
        \end{subfigure}
        \hfill
        \begin{subfigure}[b]{0.475\textwidth}   
            \centering 
            \includegraphics[width=\textwidth]{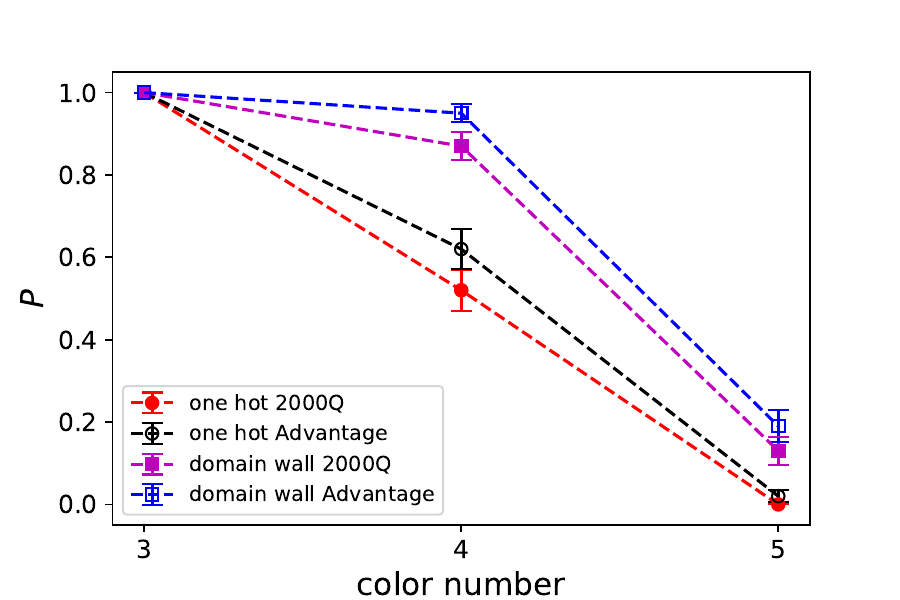}
            \caption[]{Success probability}    
            \label{fig:kcol_success}
        \end{subfigure}
        \caption[k-coloring Problem]
        {Mean of the four performance measures against the problem size for the k-coloring problem for various numbers of colours and comparing all four QPU-encoding combinations. The cost function in part (c) has been normalized by edge number to allow a more direct comparisons at different sizes. The bars for parts (a-c) are the standard deviation of the distribution, rather than standard error. Since all data points are based on $100$ samples, the standard error is ten times smaller than what is depicted by these bars. Bars for (d) are standard error. Note that the x-axis values are not the same on each graph, this is due to the fact that for subfigure (c) no QPU encoding combination returned valid solutions for all problems, and for subfigure (d) our exhaustive method of finding solutions failed beyond size $5$, and we felt the trend was clear enough from these points.} 
        \label{fig:kcol}
\end{figure*}

\subsection{$k$-Coloring}

Now that we have demonstrated that the domain-wall encoding leads to significantly better performance in encoding discrete variables with both two and three possible values, the next natural question is what happens at higher values, particularly because \cite{chancellor2019} found that these were the cases where the structure had the most effect on embedding efficiency.  To do this we examine maximum $k$-coloring problems for which the value of $k$ scales with the number of nodes. Figure~\ref{fig:kcol} shows the results. We see a similar pattern as before, with Figure~\ref{fig:kcol_chain} showing that the QPU structure makes a bigger difference in terms of chain breaks, however, at least for the Advantage QPU, usage of the domain-wall encoding can also significantly reduce the number of breaks. Also as seen in the three-Coloring case, the encoding type is the dominant factor in determining the number of solutions which are decoded correctly. Unlike the three-coloring case however, $R_{\mathrm{enc}}$ decreases toward zero as the number of colors and therefore the problem size, is increased. This is likely because for more colors, there are more possible ways to violate the constraint. The final cost function likewise shows encoding being the dominant factor in determining  performance and being a more significant factor than QPU type. We further see this trend in probability of finding the most optimal solution, although the performance difference between the Advantage and 2000Q by this metric is larger than the three-Coloring case, suggesting that the slightly better performance from Advantage is a real effect, as we will see later, a different method of comparison actually favors the older version of the processor, the 2000Q.

\begin{table*}[t]
\begin{tabular}{|c|c|c|c|c|c|c|c|c|c|c|c|c|} 
 \hline& \multicolumn{2}{c|}{Adv. dw/oh} & \multicolumn{2}{c|}{2000Q dw/oh} & \multicolumn{2}{c|}{dw Adv./2000Q} & \multicolumn{2}{c|}{oh Adv./2000Q} & \multicolumn{2}{c|}{(dw, Adv.)/(oh, 2000Q)} & \multicolumn{2}{c|}{(dw, 2000Q)/(oh, Adv.)}\tabularnewline 
 \hline 
 \hline 
3 color (b,w) & \hspace{0.2 cm} 0 \hspace{0.2 cm} & 0 &  \hspace{0.2 cm} 0 \hspace{0.2 cm}     & 0 &  \hspace{0.2 cm} 0 \hspace{0.2 cm}  & 0 &  \hspace{0.2 cm} 0 \hspace{0.2 cm}  & 0 &  \hspace{0.2 cm} 0 \hspace{0.2 cm}  & 0 &  \hspace{0.2 cm} 0 \hspace{0.2 cm}  & 0 \tabularnewline 
 \hline 
3 color  p & \multicolumn{2}{c|}{ } & \multicolumn{2}{c|}{ } & \multicolumn{2}{c|}{ } & \multicolumn{2}{c|}{ } & \multicolumn{2}{c|}{ } & \multicolumn{2}{c|}{ }  \tabularnewline 
 \hline 
4 color (b,w) & \hspace{0.2 cm} 34 \hspace{0.2 cm} & 1 &  \hspace{0.2 cm} 37 \hspace{0.2 cm}     & 2 &  \hspace{0.2 cm} 11 \hspace{0.2 cm}  & 3 &  \hspace{0.2 cm} 26 \hspace{0.2 cm}  & 16 &  \hspace{0.2 cm} 44 \hspace{0.2 cm}  & 1 &  \hspace{0.2 cm} 33 \hspace{0.2 cm}  & 7 \tabularnewline 
 \hline 
4 color  p & \multicolumn{2}{c|}{$1.05\times 10^{-9}$\cellcolor{green!50}} & \multicolumn{2}{c|}{$1.42\times 10^{-9}$\cellcolor{green!50}} & \multicolumn{2}{c|}{$2.87\times 10^{-2}$\cellcolor{green!50}} & \multicolumn{2}{c|}{$8.21\times 10^{-2}$\cellcolor{yellow!50}} & \multicolumn{2}{c|}{$1.31\times 10^{-12}$\cellcolor{green!50}} & \multicolumn{2}{c|}{$2.11\times 10^{-5}$\cellcolor{green!50}}  \tabularnewline 
 \hline 
5 color (b,w) & \hspace{0.2 cm} 91 \hspace{0.2 cm} & 1 &  \hspace{0.2 cm} 78 \hspace{0.2 cm}     & 1 &  \hspace{0.2 cm} 34 \hspace{0.2 cm}  & 18 &  \hspace{0.2 cm} 23 \hspace{0.2 cm}  & 59 &  \hspace{0.2 cm} 88 \hspace{0.2 cm}  & 1 &  \hspace{0.2 cm} 91 \hspace{0.2 cm}  & 1 \tabularnewline 
 \hline 
5 color  p & \multicolumn{2}{c|}{$1.88\times 10^{-26}$\cellcolor{green!50}} & \multicolumn{2}{c|}{$1.32\times 10^{-22}$\cellcolor{green!50}} & \multicolumn{2}{c|}{$1.82\times 10^{-2}$\cellcolor{green!50}} & \multicolumn{2}{c|}{$ \approx 1 $\cellcolor{red!50}} & \multicolumn{2}{c|}{$1.45\times 10^{-25}$\cellcolor{green!50}} & \multicolumn{2}{c|}{$1.88\times 10^{-26}$\cellcolor{green!50}}  \tabularnewline 
 \hline 
6 color(b,w) & \hspace{0.2 cm} 99 \hspace{0.2 cm} & 0 &  \hspace{0.2 cm}   &  FAIL &  \hspace{0.2 cm} 59 \hspace{0.2 cm}  & 15 &  \hspace{0.2 cm}   &  FAIL &  \hspace{0.2 cm}   &  FAIL &  \hspace{0.2 cm} 99 \hspace{0.2 cm}  & 0 \tabularnewline 
 \hline 
6 color p & \multicolumn{2}{c|}{$1.58\times 10^{-30}$\cellcolor{green!50}} & \multicolumn{2}{c|}{ } & \multicolumn{2}{c|}{$1.28\times 10^{-7}$\cellcolor{green!50}} & \multicolumn{2}{c|}{ } & \multicolumn{2}{c|}{ } & \multicolumn{2}{c|}{$1.58\times 10^{-30}$\cellcolor{green!50}}  \tabularnewline 
 \hline 
7 color(b,w) & \hspace{0.2 cm} 92 \hspace{0.2 cm} & 0 &  FAIL &  FAIL &  \hspace{0.2 cm}   &  FAIL &  \hspace{0.2 cm}   &  FAIL &  \hspace{0.2 cm}   &  FAIL &  FAIL &   \tabularnewline 
 \hline 
7 color p & \multicolumn{2}{c|}{$2.02\times 10^{-28}$\cellcolor{green!50}} & \multicolumn{2}{c|}{ } & \multicolumn{2}{c|}{ } & \multicolumn{2}{c|}{ } & \multicolumn{2}{c|}{ } & \multicolumn{2}{c|}{ }  \tabularnewline 
 \hline 
\end{tabular} 
 
\caption{\label{tab:hypo_k} Hypothesis testing results for all six possible comparisons of QPU-encoding combinations for k color problems with different numbers of colors. 
For each comparison the expected winner is listed first. 
For each size the count of cases where the expected winner (written first at the top of the column) performs better $n_b$ (left) and worse $n_w$ (right) are listed. 
Below is listed the value of $p$ as calculated by Equation~\eqref{eq:signifcance}. 
In cases where either both combinations perform the same on all problems, or one or both fail to embed, statistical significance cannot be calculated. 
In case where the expected winner failed to embed, we write `FAIL' in the left column, and likewise if the embedding fails for the QPU-encoding combination described by the right column. 
These comparisons are performed for the single best solution found out of all $100$ samples, using majority vote decoding for broken chains. 
If none of the samples decode to valid solutions, than the cost function is treated as being ``infinite'' and any finite value is considered to be better. 
Color coding used as a guide to the eye, green indicates a statistically significant rejection of the null hypothesis, while yellow indicates a result which is not statistically significant. 
Red indicates a statistically significant result which rejects the alternate hypothesis.}
\end{table*}

\subsection{Hypothesis Testing for $k$-Coloring}
As we have done before, we perform hypothesis testing based analysis on the best cost function values returned by all QPU-encoding combinations. 
As with the three-Coloring case, the domain-wall encoding leads to a statistically significant improvement at all but the smallest size. 
Even when we perform a cross comparison, we see that the domain-wall encoding on a 2000Q out-performs the one-hot encoding on the Advantage. 
We do find one surprising result, not in terms of comparison between the two encodings, but in terms of comparisons between the two QPUs. 
For the one-hot encoding of the maximum five coloring problem, we find a highly statistically significant result that the 2000Q actually outperforms the Advantage processor. 
This is the opposite trend to what is seen when the domain-wall encoding is used, and very unusual since a more highly connected device should perform better than a less connected one. 
To understand the root cause of this effect, we perform the same analysis, but discard broken-chain solutions, rather than perform majority vote decoding. 
As Table~\ref{tab:hypo_k_noDec} in the appendix shows, the effect goes away when we change decoding strategy, indicating that this is an artifact of the strategy. Since the primary purpose of this paper is to compare encoding strategy, rather than QPU performance, we have elected not to probe this effect further, although doing so could potentially yield interesting results

 \section{Discussion and conclusions}
We have performed the first (to our knowledge) experimental tests of the domain-wall encoding proposed in \cite{chancellor2019} on quantum annealing processors. We find that for problems with variables up to size seven, the domain-wall encoding out-performs the traditional one-hot encoding in all but the smallest cases, for which their performance is roughly equal due to the ease of the problems. We further find that the domain-wall encoding generally reduces the number of broken minor-embedding chains and increases the proportion of solutions which decode correctly. Crucially, for every problem we look at, we do not find a single metric for which the domain-wall encoding performs worse on average than one-hot on the same QPU, suggesting that the domain-wall encoding should be the method of choice.

Dramatically, when we perform a cross comparison of performance between domain-wall encoding on the older 2000Q QPU versus one-hot on the newer Advantage QPU, we find that the encoding makes a bigger difference in solution quality than the QPU architecture (at least on problems which can be embedded into both QPUs, embedding fails at a smaller size for domain wall on a 2000Q than it does for one hot on Advantage). This underscores the importance of ``software'' advances like better encoding in parallel to hardware advances. This is particularly true given that, while developing new hardware can be a very expensive endeavor, using a different encoding can be done at almost no cost. Given the recent trend for the programming of these devices to be done at an increasingly higher level, the end user does not necessarily need to even ``see'' the encoding steps at all, for example the D-Wave ocean repository currently has a discrete quadratic model (DQM) solver \cite{DQM}, which uses one-hot encoding to solve optimization problems \cite{DQM_forum}. Changing the underlying encoding used by this solver is likely to improve performance, but would have no other effect on the way the end users interact with this solver, and would not require them to understand how this strategy works. 
The German Aerospace Center (DLR) is developing a library which is more hardware agnostic (including gate-based quantum computers) with overlapping functionality.
It is planed to integrate different encodings, including domain-wall encoding, and publish the software under an open-source license.

Although not the focus of the present study, it would be very illuminating to examine the underlying cause of the drop in success probability as size increases. In particular, it has been observed for example in \cite{chancellor20b} that even for an isolated domain wall chain without a programmed potential, the probability of finding a domain wall at different locations is far from uniform, especially if spin reversal transforms are not performed. In that paper the underlying cause was analog noise on the device causing biases toward some configurations, these biases may have been enhanced by the fact that the dynamics of one dimensional chains tend to freeze very late in the anneal \cite{Izquierdo20a}.

While it being not the main purpose of our study, we have also compared the Advantage and 2000Q QPUs. We have found that minor embedding chains break less frequently on the Advantage QPU, as what should be expected on a more connected graph. We found that the Advantage QPU also performed better (or for small problems no statistically significant difference could be found) at solving problems in all but one example on a five coloring problem. This result is highly statistically significant, so it is unlikely to be a statistical anomaly. And it also goes away when we do not perform majority vote decoding on the broken chains. While we have not investigated this effect further since it is far from our main purpose, it is likely that a more complete investigation could be fruitful in finding improved broken-chain decoding strategies. It is also worth remembering that there was room for improvement in how the constraint strength was chosen for Advantage, a more optimal strategy here would change the comparison between Advantage and 2000Q (further) in favor of Advantage.

While our work gives compelling evidence that the domain-wall-strategy is superior for currently available superconducting flux qubit QPUs, there are still many unanswered questions with regards to the encoding which are beyond the scope of our current work. 
It would be illuminating to test these strategies on larger problems and more connected hardware graphs using quantum Monte Carlo. This approach has previously been used to compare embedding strategies \cite{Albash16b}. 
It could further be interesting to test whether the advantages seen in annealing carry over to gate model optimization algorithms.
Preliminary work \cite{chancellor2019} gives some theoretical suggestions for this, because our the domain-wall encoding approach will require fewer interactions between distant qubits. 
However, an experimental test would be enlightening.
 
\section{Acknowledgments}

NC and JC were funded by UKRI EPSRC grant number EP/S00114X/1 and QPU access for early experiments was supported by impact acceleration funding associated with grant EP/L022303/1, although none of these data are directly presented in the manuscript. 
The authors gratefully acknowledge the J\"ulich Supercomputing Centre~\footnote{\url{https://www.fz-juelich.de/ias/jsc}}) for funding this project by providing computing time through the J\"ulich UNified Infrastructure for Quantum computing (JUNIQ) an the D-Wave quantum  annealer.
 
\section*{Appendix: hypothesis testing tables discarding broken chains}

In this appendix we provide versions of Tables~\ref{tab:hypo_three} and~\ref{tab:hypo_k} but where solutions with broken chains are discarded rather than decoded by majority vote, the results appear in Tables~\ref{tab:hypo_three_noDec} and~\ref{tab:hypo_k_noDec} respectively.

\begin{table*}[t]

\begin{tabular}{|c|c|c|c|c|c|c|c|c|c|c|c|c|} 
 \hline& \multicolumn{2}{c|}{Adv. dw/oh} & \multicolumn{2}{c|}{2000Q dw/oh} & \multicolumn{2}{c|}{dw Adv./2000Q} & \multicolumn{2}{c|}{oh Adv./2000Q} & \multicolumn{2}{c|}{(dw, Adv.)/(oh, 2000Q)} & \multicolumn{2}{c|}{(dw, 2000Q)/(oh, Adv.)}\tabularnewline 
 \hline 
 \hline 
5 node (b,w) & \hspace{0.2 cm} 0 \hspace{0.2 cm} & 0 &  \hspace{0.2 cm} 0 \hspace{0.2 cm}     & 0 &  \hspace{0.2 cm} 0 \hspace{0.2 cm}  & 0 &  \hspace{0.2 cm} 0 \hspace{0.2 cm}  & 0 &  \hspace{0.2 cm} 0 \hspace{0.2 cm}  & 0 &  \hspace{0.2 cm} 0 \hspace{0.2 cm}  & 0 \tabularnewline 
 \hline 
5 node  p & \multicolumn{2}{c|}{ } & \multicolumn{2}{c|}{ } & \multicolumn{2}{c|}{ } & \multicolumn{2}{c|}{ } & \multicolumn{2}{c|}{ } & \multicolumn{2}{c|}{ }  \tabularnewline 
 \hline 
10 node (b,w) & \hspace{0.2 cm} 42 \hspace{0.2 cm} & 0 &  \hspace{0.2 cm} 38 \hspace{0.2 cm}     & 0 &  \hspace{0.2 cm} 2 \hspace{0.2 cm}  & 0 &  \hspace{0.2 cm} 22 \hspace{0.2 cm}  & 21 &  \hspace{0.2 cm} 40 \hspace{0.2 cm}  & 0 &  \hspace{0.2 cm} 40 \hspace{0.2 cm}  & 0 \tabularnewline 
 \hline 
10 node  p & \multicolumn{2}{c|}{$2.27\times 10^{-13}$\cellcolor{green!50}} & \multicolumn{2}{c|}{$3.64\times 10^{-12}$\cellcolor{green!50}} & \multicolumn{2}{c|}{$2.50\times 10^{-1}$\cellcolor{yellow!50}} & \multicolumn{2}{c|}{$5.00\times 10^{-1}$\cellcolor{yellow!50}} & \multicolumn{2}{c|}{$9.09\times 10^{-13}$\cellcolor{green!50}} & \multicolumn{2}{c|}{$9.09\times 10^{-13}$\cellcolor{green!50}}  \tabularnewline 
 \hline 
15 node (b,w) & \hspace{0.2 cm} 88 \hspace{0.2 cm} & 2 &  \hspace{0.2 cm} 96 \hspace{0.2 cm}     & 3 &  \hspace{0.2 cm} 35 \hspace{0.2 cm}  & 28 &  \hspace{0.2 cm} 78 \hspace{0.2 cm}  & 16 &  \hspace{0.2 cm} 96 \hspace{0.2 cm}  & 0 &  \hspace{0.2 cm} 91 \hspace{0.2 cm}  & 3 \tabularnewline 
 \hline 
15 node  p & \multicolumn{2}{c|}{$3.31\times 10^{-24}$\cellcolor{green!50}} & \multicolumn{2}{c|}{$2.55\times 10^{-25}$\cellcolor{green!50}} & \multicolumn{2}{c|}{$2.25\times 10^{-1}$\cellcolor{yellow!50}} & \multicolumn{2}{c|}{$2.89\times 10^{-11}$\cellcolor{green!50}} & \multicolumn{2}{c|}{$1.26\times 10^{-29}$\cellcolor{green!50}} & \multicolumn{2}{c|}{$6.99\times 10^{-24}$\cellcolor{green!50}}  \tabularnewline 
 \hline 
20 node (b,w) & \hspace{0.2 cm} 99 \hspace{0.2 cm} & 1 &  \hspace{0.2 cm} 96 \hspace{0.2 cm}     & 0 &  \hspace{0.2 cm} 60 \hspace{0.2 cm}  & 33 &  \hspace{0.2 cm} 98 \hspace{0.2 cm}  & 0 &  \hspace{0.2 cm} 100 \hspace{0.2 cm}  & 0 &  \hspace{0.2 cm} 82 \hspace{0.2 cm}  & 15 \tabularnewline 
 \hline 
20 node  p & \multicolumn{2}{c|}{$7.97\times 10^{-29}$\cellcolor{green!50}} & \multicolumn{2}{c|}{$1.26\times 10^{-29}$\cellcolor{green!50}} & \multicolumn{2}{c|}{$3.35\times 10^{-3}$\cellcolor{green!50}} & \multicolumn{2}{c|}{$3.16\times 10^{-30}$\cellcolor{green!50}} & \multicolumn{2}{c|}{$7.89\times 10^{-31}$\cellcolor{green!50}} & \multicolumn{2}{c|}{$1.19\times 10^{-12}$\cellcolor{green!50}}  \tabularnewline 
 \hline 
25 node (b,w) & \hspace{0.2 cm} 100 \hspace{0.2 cm} & 0 &  \hspace{0.2 cm}   &  FAIL &  \hspace{0.2 cm} 93 \hspace{0.2 cm}  & 6 &  \hspace{0.2 cm}   &  FAIL &  \hspace{0.2 cm}   &  FAIL &  \hspace{0.2 cm} 61 \hspace{0.2 cm}  & 24 \tabularnewline 
 \hline 
25 node  p & \multicolumn{2}{c|}{$7.89\times 10^{-31}$\cellcolor{green!50}} & \multicolumn{2}{c|}{ } & \multicolumn{2}{c|}{$1.89\times 10^{-21}$\cellcolor{green!50}} & \multicolumn{2}{c|}{ } & \multicolumn{2}{c|}{ } & \multicolumn{2}{c|}{$3.70\times 10^{-5}$\cellcolor{green!50}}  \tabularnewline 
 \hline 
30 node (b,w) & \hspace{0.2 cm} 100 \hspace{0.2 cm} & 0 &  \hspace{0.2 cm}   &  FAIL &  \hspace{0.2 cm} 100 \hspace{0.2 cm}  & 0 &  \hspace{0.2 cm}   &  FAIL &  \hspace{0.2 cm}   &  FAIL &  \hspace{0.2 cm} 14 \hspace{0.2 cm}  & 3 \tabularnewline 
 \hline 
30 node  p & \multicolumn{2}{c|}{$7.89\times 10^{-31}$\cellcolor{green!50}} & \multicolumn{2}{c|}{ } & \multicolumn{2}{c|}{$7.89\times 10^{-31}$\cellcolor{green!50}} & \multicolumn{2}{c|}{ } & \multicolumn{2}{c|}{ } & \multicolumn{2}{c|}{$6.36\times 10^{-3}$\cellcolor{green!50}}  \tabularnewline 
 \hline 
35 node (b,w) & \hspace{0.2 cm} 100 \hspace{0.2 cm} & 0 &  FAIL &  FAIL &  \hspace{0.2 cm}   &  FAIL &  \hspace{0.2 cm}   &  FAIL &  \hspace{0.2 cm}   &  FAIL &  FAIL &   \tabularnewline 
 \hline 
35 node  p & \multicolumn{2}{c|}{$7.89\times 10^{-31}$\cellcolor{green!50}} & \multicolumn{2}{c|}{ } & \multicolumn{2}{c|}{ } & \multicolumn{2}{c|}{ } & \multicolumn{2}{c|}{ } & \multicolumn{2}{c|}{ }  \tabularnewline 
 \hline 
40 node(b,w) & \hspace{0.2 cm} 88 \hspace{0.2 cm} & 0 &  FAIL &  FAIL &  \hspace{0.2 cm}   &  FAIL &  \hspace{0.2 cm}   &  FAIL &  \hspace{0.2 cm}   &  FAIL &  FAIL &   \tabularnewline 
 \hline 
40 node p & \multicolumn{2}{c|}{$3.23\times 10^{-27}$\cellcolor{green!50}} & \multicolumn{2}{c|}{ } & \multicolumn{2}{c|}{ } & \multicolumn{2}{c|}{ } & \multicolumn{2}{c|}{ } & \multicolumn{2}{c|}{ }  \tabularnewline 
 \hline 
\end{tabular} 
 
\caption{\label{tab:hypo_three_noDec}Hypothesis testing results for all six possible comparisons of QPU-encoding combinations for three color problems of different sizes. For each comparison the expected winner is listed first. For each size the count of cases where the expected winner  performs better $n_b$ (left) and worse $n_w$ (right) are listed. Below is listed the value of $p$ as calculated by Equation~\eqref{eq:signifcance}. In cases where either both combinations perform the same on all problems, or one or both fail to embed, statistical significance cannot be calculated. In case where the expected winner failed to embed, we write `FAIL' in the left column, and likewise if the embedding fails for the QPU-encoding combination described by the right column. These comparisons are performed for the single best solution found out of all $100$ samples, where samples with broken chains are treated as being invalid. If none of the samples decode to valid solutions, than the cost function is treated as being ``infinite'' and any finite value is considered to be better. Color coding used as a guide to the eye, green indicates a statistically significant rejection of the null hypothesis, while yellow indicates a result which is not statistically significant.}
\end{table*}

\begin{table*}[t]

\begin{tabular}{|c|c|c|c|c|c|c|c|c|c|c|c|c|} 
 \hline& \multicolumn{2}{c|}{Adv. dw/oh} & \multicolumn{2}{c|}{2000Q dw/oh} & \multicolumn{2}{c|}{dw Adv./2000Q} & \multicolumn{2}{c|}{oh Adv./2000Q} & \multicolumn{2}{c|}{(dw, Adv.)/(oh, 2000Q)} & \multicolumn{2}{c|}{(dw, 2000Q)/(oh, Adv.)}\tabularnewline 
 \hline 
 \hline 
3 color (b,w) & \hspace{0.2 cm} 0 \hspace{0.2 cm} & 0 &  \hspace{0.2 cm} 0 \hspace{0.2 cm}     & 0 &  \hspace{0.2 cm} 0 \hspace{0.2 cm}  & 0 &  \hspace{0.2 cm} 0 \hspace{0.2 cm}  & 0 &  \hspace{0.2 cm} 0 \hspace{0.2 cm}  & 0 &  \hspace{0.2 cm} 0 \hspace{0.2 cm}  & 0 \tabularnewline 
 \hline 
3 color  p & \multicolumn{2}{c|}{ } & \multicolumn{2}{c|}{ } & \multicolumn{2}{c|}{ } & \multicolumn{2}{c|}{ } & \multicolumn{2}{c|}{ } & \multicolumn{2}{c|}{ }  \tabularnewline 
 \hline 
4 color (b,w) & \hspace{0.2 cm} 39 \hspace{0.2 cm} & 1 &  \hspace{0.2 cm} 42 \hspace{0.2 cm}     & 3 &  \hspace{0.2 cm} 17 \hspace{0.2 cm}  & 2 &  \hspace{0.2 cm} 33 \hspace{0.2 cm}  & 14 &  \hspace{0.2 cm} 53 \hspace{0.2 cm}  & 1 &  \hspace{0.2 cm} 32 \hspace{0.2 cm}  & 9 \tabularnewline 
 \hline 
4 color  p & \multicolumn{2}{c|}{$3.73\times 10^{-11}$\cellcolor{green!50}} & \multicolumn{2}{c|}{$4.33\times 10^{-10}$\cellcolor{green!50}} & \multicolumn{2}{c|}{$3.64\times 10^{-4}$\cellcolor{green!50}} & \multicolumn{2}{c|}{$3.97\times 10^{-3}$\cellcolor{green!50}} & \multicolumn{2}{c|}{$3.05\times 10^{-15}$\cellcolor{green!50}} & \multicolumn{2}{c|}{$2.15\times 10^{-4}$\cellcolor{green!50}}  \tabularnewline 
 \hline 
5 color (b,w) & \hspace{0.2 cm} 97 \hspace{0.2 cm} & 1 &  \hspace{0.2 cm} 89 \hspace{0.2 cm}     & 2 &  \hspace{0.2 cm} 66 \hspace{0.2 cm}  & 7 &  \hspace{0.2 cm} 70 \hspace{0.2 cm}  & 7 &  \hspace{0.2 cm} 100 \hspace{0.2 cm}  & 0 &  \hspace{0.2 cm} 80 \hspace{0.2 cm}  & 9 \tabularnewline 
 \hline 
5 color  p & \multicolumn{2}{c|}{$3.12\times 10^{-28}$\cellcolor{green!50}} & \multicolumn{2}{c|}{$1.69\times 10^{-24}$\cellcolor{green!50}} & \multicolumn{2}{c|}{$1.92\times 10^{-13}$\cellcolor{green!50}} & \multicolumn{2}{c|}{$1.76\times 10^{-14}$\cellcolor{green!50}} & \multicolumn{2}{c|}{$7.89\times 10^{-31}$\cellcolor{green!50}} & \multicolumn{2}{c|}{$1.15\times 10^{-15}$\cellcolor{green!50}}  \tabularnewline 
 \hline 
6 color(b,w) & \hspace{0.2 cm} 100 \hspace{0.2 cm} & 0 &  \hspace{0.2 cm}   &  FAIL &  \hspace{0.2 cm} 98 \hspace{0.2 cm}  & 0 &  \hspace{0.2 cm}   &  FAIL &  \hspace{0.2 cm}   &  FAIL &  \hspace{0.2 cm} 26 \hspace{0.2 cm}  & 0 \tabularnewline 
 \hline 
6 color p & \multicolumn{2}{c|}{$7.89\times 10^{-31}$\cellcolor{green!50}} & \multicolumn{2}{c|}{ } & \multicolumn{2}{c|}{$3.16\times 10^{-30}$\cellcolor{green!50}} & \multicolumn{2}{c|}{ } & \multicolumn{2}{c|}{ } & \multicolumn{2}{c|}{$1.49\times 10^{-8}$\cellcolor{green!50}}  \tabularnewline 
 \hline 
7 color(b,w) & \hspace{0.2 cm} 38 \hspace{0.2 cm} & 0 &  FAIL &  FAIL &  \hspace{0.2 cm}   &  FAIL &  \hspace{0.2 cm}   &  FAIL &  \hspace{0.2 cm}   &  FAIL &  FAIL &   \tabularnewline 
 \hline 
7 color p & \multicolumn{2}{c|}{$3.64\times 10^{-12}$\cellcolor{green!50}} & \multicolumn{2}{c|}{ } & \multicolumn{2}{c|}{ } & \multicolumn{2}{c|}{ } & \multicolumn{2}{c|}{ } & \multicolumn{2}{c|}{ }  \tabularnewline 
 \hline 
\end{tabular} 
 
\caption{\label{tab:hypo_k_noDec}Hypothesis testing results for all six possible comparisons of QPU-encoding combinations for k-color problems of different number of colors. For each comparison the expected winner is listed first. For each size the count of cases where the expected winner performs better $n_b$ (left) and worse $n_w$ (right) are listed. Below is listed the value of $p$ as calculated by Equation~\eqref{eq:signifcance}. In cases where either both combinations perform the same on all problems, or one or both fail to embed, statistical significance cannot be calculated. In case where the expected winner failed to embed, we write `FAIL' in the left column, and likewise if the embedding fails for the QPU-encoding combination described by the right column. These comparisons are performed for the single best solution found out of all $100$ samples, where samples with broken chains are treated as being invalid. If none of the samples decode to valid solutions, than the cost function is treated as being ``infinite'' and any finite value is considered to be better. Color coding used as a guide to the eye, green indicates a statistically significant rejection of the null hypothesis.}
\end{table*}
 
\bibliography{references}

\end{document}